\documentclass[manuscript,screen,nonacm]{acmart}

\setcopyright{none}

\AtBeginDocument{
  }

\usepackage{graphicx}
\usepackage{subcaption}
\usepackage{makecell}
\usepackage{threeparttable}
\usepackage{longtable}
\usepackage{array}
\usepackage{multirow}
\usepackage{booktabs}

\usepackage{placeins}

\begin{document}

\title{Learning When to Intervene on Habitual Behaviors: A Case Study in Oral Health Care}

\author{Bhanu Teja Gullapalli}
\email{bgullapalli@g.harvard.edu}
\affiliation{
  \department{Department of Statistics and School of Engineering and Applied Sciences}
  \institution{Harvard University}
  \city{Cambridge}
  \state{Massachusetts}
  \country{USA}
}

\author{Vivek Shetty}
\email{vshetty@g.ucla.edu}
\affiliation{
  \department{School of Dentistry}
  \institution{University of California, Los Angeles}
  \city{Los Angeles}
  \state{California}
  \country{USA}
}

\author{Anna L. Trella}
\email{atrella@g.harvard.edu}
\affiliation{
  \department{Department of Statistics and School of Engineering and Applied Sciences}
  \institution{Harvard University}
  \city{Cambridge}
  \state{Massachusetts}
  \country{USA}
}

\author{Asim H. Gazi}
\email{agazi@seas.harvard.edu}
\affiliation{
  \department{Department of Statistics and School of Engineering and Applied Sciences}
  \institution{Harvard University}
  \city{Cambridge}
  \state{Massachusetts}
  \country{USA}
}

\author{Susan A. Murphy}
\email{samurphy@g.harvard.edu}
\authornote{Susan A. Murphy holds concurrent appointments at Harvard University and as an Amazon Scholar. This paper describes work performed at Harvard University and is not associated with Amazon.}
\affiliation{
  \department{Department of Statistics and School of Engineering and Applied Sciences}
  \institution{Harvard University}
  \city{Cambridge}
  \state{Massachusetts}
  \country{USA}
}

\renewcommand{\shortauthors}{Gullapalli et al.}

\begin{abstract}
A central challenge for digital health interventions aimed at improving habitual behaviors is deciding when to deliver an intervention prompt. For many daily habits—such as tooth brushing or eating—individuals tend to act around a usual time of day, but this timing is not fixed and can shift as routines evolve. When intervention timing is selected in advance and held constant throughout a study, it can gradually become misaligned with behavior, causing interventions to potentially arrive after the behavior has already occurred or too early to be effective. In this work, we address this habitual timing misalignment in digital health interventions by proposing an online decision-making framework that continuously adapts intervention timing as individual behavior patterns change.
Rather than treating intervention timing as a static design choice, our framework adapts it over time and integrates it into a sequential process that determines both \textit{when} and \textit{whether} to deliver an intervention. Using data from a deployed oral health intervention trial as a case study, we evaluate our approach using both observed data and simulated settings to assess how well different intervention timing strategies align with the timing of brushing events. Across these evaluations, we measure performance using a coverage-based metric that captures whether an intervention is delivered sufficiently close to a subsequent brushing event. We find that adaptive intervention timing consistently improves coverage compared to fixed intervention times based on user-provided input. The proposed framework is currently deployed in an
ongoing randomized controlled trial of a digital oral health intervention, with preliminary results that are consistent with and further support our prior evaluations.
\end{abstract}



 \maketitle

\section{Introduction }

Digital health interventions aim to support health-related behaviors by delivering timely and relevant prompts in individual's everyday lives. In recent years, the widespread adoption of smartphones, wearable devices, and connected health technologies has enabled the continuous collection of fine-grained behavioral and contextual data outside of clinical settings. These technologies have made it increasingly feasible to monitor daily activities such as physical activity, eating, sleep, or oral self-care in real time, and to deliver interventions directly through personal devices as individuals go about their routines \cite{liao2020personalized, chen2020promotion, pape2025effects, trella2024oralytics}.

As digital health systems increasingly interact with individuals over extended periods of time, they must make repeated decisions about how and when to intervene in the context of daily life. In these longitudinal settings, the timing of an intervention becomes a central challenge. This challenge is particularly pronounced for habitual behaviors—such as tooth brushing, eating, or physical activity—which tend to occur around a usual time of day but can shift as routines evolve. In many such applications, intervention prompts are intended to influence behavior \emph{before} it occurs, requiring systems not only to anticipate when an individual is most likely to engage in the target behavior, but also to proactively schedule intervention delivery at an appropriate time. For example, an intervention to improve brushing behavior is most useful if delivered shortly before brushing occurs; if it arrives after brushing, it provides no benefit, and if it arrives too early, it may be perceived as irrelevant or forgotten. Over time, even small deviations between an individual’s actual behavior and the time at which interventions are delivered can accumulate, leading to persistent misalignment between intervention delivery and behavior. Such misalignment primarily affects the relevance and usefulness of interventions, rather than whether the behavior occurs at all.

Despite the importance of intervention timing, many digital health interventions for habitual behaviors treat timing as a static design choice. Common approaches rely on user-specified preferences, clinician-defined heuristics, or population-level defaults that are selected at the start and held fixed throughout the study \cite{nahum2016just, walton2018optimizing, klasnja2015microrandomized}. These approaches implicitly assume that the timing of an individual’s routine remains stable over time, and do not adapt as behavior shifts. There is also a substantial body of work in which intervention timing is not fixed, but instead determined by detecting or inferring an underlying state, such as stress, craving, or lapse risk \cite{battalio2021sense2stop, goldstein2020refining}. In these systems, interventions are triggered once a relevant state has been detected. More broadly, such approaches are reactive, relying on the detection of either an internal state or the behavior itself to initiate intervention delivery. Prior work in digital health has demonstrated the feasibility of detecting a wide range of health-related behaviors using data from wearable and mobile sensors \cite{thomaz2015practical, yuan2024self, gullapalli2021opitrack, huang2025ibrush}. However, these reactive paradigms are insufficient for habitual behaviors where interventions must be scheduled in advance. In such settings, intervention timing decisions must be made under uncertainty, based on anticipated behavior rather than detected events.

Intervention timing is therefore an integral part of the intervention system. Each time a system makes a decision about intervention delivery, it must determine not only whether an intervention should be delivered, but also when it should be delivered relative to an individual’s ongoing behavior. In this work, we treat intervention timing and intervention delivery as two distinct but coupled decisions. Specifically, we focus on settings in which the timing of an intervention must be selected in advance of when the target behavior may occur. At each opportunity to intervene, we first determine a candidate intervention time using an individual-specific model learned from longitudinal behavioral data. We then determine whether an intervention should be delivered at that time using an intervention delivery policy,  which specifies the probability or rule governing whether a prompt is sent at the candidate intervention time.

In this paper, we make the following key contributions:
\begin{itemize}

\item \textbf{\emph{We identify an underexplored intervention timing challenge in longitudinal digital health interventions.}}
We highlight settings in which intervention times must be chosen in advance of a habitual behavior and cannot be determined by detecting the behavior itself. In such settings, fixed or user-provided schedules can gradually become misaligned as routines change, even when intervention delivery and content remain unchanged.

    \item \textbf{\emph{We empirically characterize intervention timing misalignment using data from a real-world digital health intervention deployment.}}
    Using data from a prior digital oral health intervention, we show how intervention times specified at start of the study can become misaligned with observed brushing behavior, revealing the prevalence and variability of timing misalignment in practice.

\item \textbf{\emph{We develop an online method for adapting intervention timing based on observed user behavior.}}

We introduce an online learning framework that continuously updates individual-specific estimates of behavior (e.g., brushing time) and associated uncertainty as new user data is observed. These estimates are used to select intervention times. To make ideas concrete, we apply this framework to the oral health setting and consider updating individual's brushing time estimates as new brushing data is observed.

\item \textbf{\emph{We evaluate adaptive intervention timing using two complementary forms of analysis.}}
Using data from a real-world oral health intervention, we (1) conduct an offline evaluation by leveraging observed brushing behavior to compare how adaptive intervention times align with behavior relative to fixed, user-defined schedules, and (2) construct simulation environments that model real-world deployment, where intervention timing decisions can influence subsequent behavior, allowing us to evaluate alignment under a full online decision-making pipeline.

\item \textbf{\emph{We deploy the proposed framework in an ongoing clinical trial and report preliminary findings.}}
The framework is currently deployed in a real-world oral health intervention, where preliminary results are consistent with those observed in our prior evaluations, providing early evidence of its effectiveness in practice.
\end{itemize}

To our knowledge, this is the first work to develop and evaluate  adaptive intervention timing methods for habitual behaviors.

\section{Related Work}

Digital health interventions aimed at supporting habitual behavior change operate as sequential decision-making systems, making repeated decisions over time about \emph{whether} to deliver an intervention, \emph{what} content to deliver, and \emph{when} to deliver it \cite{nahum2016just}. Intervention timing in such systems can be determined in two broad ways. In some settings, interventions must be scheduled in advance based on predictions of when a behavior is likely to occur, requiring systems to anticipate future behavior under uncertainty. In other settings, interventions are triggered reactively in response to detected behaviors or inferred internal states. These two paradigms—anticipatory and reactive timing—reflect different assumptions about when and how intervention decisions can be made.

In anticipatory settings, where intervention timing must be determined in advance of the target behavior, a common approach is to rely on fixed schedules based on user-provided or predefined routines. This is common in applications such as medication adherence, where reminders are delivered at or shortly before scheduled dosing times, often tied to routine events such as meals \cite{vervloet2012effectiveness, thakkar2016mobile}. Similarly, in oral health and sleep interventions, prompts are typically scheduled based on user-reported routines and delivered at fixed offsets prior to the expected behavior. For example, oral health interventions often rely on individuals' reported brushing times (e.g., morning and evening routines) and deliver reminders shortly before these times to encourage hygiene or improve brushing quality \cite{trella2024oralytics,scheerman2020effect}. In sleep interventions, guidance and prompts are typically aligned with prescribed sleep schedules, such as target bedtimes and wake-up times, which are determined in advance based on individual routines or clinical recommendations \cite{fullagar2016time,ong2023randomized}.

In reactive settings, intervention timing is determined based on detected behaviors or inferred internal states. For example, systems such as Sense2Stop \cite{battalio2021sense2stop} deliver interventions when states such as stress or craving are inferred from sensor data, with the goal of providing timely support during moments when individuals are at increased risk of engaging in smoking \cite{naughton2017delivering,businelle2016ecological}. Other examples include interventions in mental health settings that deliver support in response to elevated stress or mood disturbances, where interventions are triggered based on inferred or self-reported affective states \cite{goldstein2020refining, ben2021smartphone}. In addition, prior work has demonstrated the feasibility of detecting behaviors such as smoking, eating, and oral self-care using wearable and mobile sensors, enabling interventions to be triggered in response to observed events \cite{parate2014risq, thomaz2015practical, bhandari2017non}. While these approaches enable responsive intervention delivery, they determine timing in reaction to detected states or behaviors. In this work, we focus on the anticipatory setting, where interventions must be delivered in advance of the behavior in order to influence it.

Existing approaches thus either rely on fixed schedules specified in advance or determine intervention timing reactively based on detected states or behaviors. However, these approaches do not address settings in which intervention timing must be selected in advance and continuously adapted as behavior patterns evolve over time. In such settings, effective intervention requires anticipating when a behavior is likely to occur and determining how far in advance to intervene to influence it. In this work, we address this gap by developing an online decision-making framework that continuously updates future intervention timings based on previously observed behavior, enabling adaptive and anticipatory intervention delivery.

\section{Digital Oral Health Intervention Dataset from a Micro-Randomized Trial}
\label{Sec:dataset}

We introduce the digital oral health intervention dataset used throughout this work. We collected this dataset as part of a previously deployed digital oral health intervention trial \footnote{ClinicalTrials.gov identifier: NCT05624489, \url{https://clinicaltrials.gov/study/NCT05624489?viewType=Card&term=Oralytics&rank=1}.}. This dataset contains time-stamped brushing events, capturing when brushing occurs, along with records of when intervention prompts were delivered during the trial. The dataset serves three primary
purposes in this paper:

\begin{itemize}

    \item It provides the empirical foundation for motivating the problem of intervention timing misalignment. We analyze the relationship between user-provided brushing times, which are used to determine intervention timing, and the timing of observed brushing events to illustrate how fixed intervention times based on self-reported inputs can become misaligned with actual brushing times over time.

    \item It is used to evaluate adaptive intervention timing against the intervention timing used in this deployment. Using observed brushing data, we compare adaptive strategies to user-provided times in terms of how well they align with, or \emph{cover}, brushing events.

    \item It is used to construct simulation environments that approximate individual brushing-time patterns. These simulations enable evaluation of adaptive intervention timing within a complete online decision-making pipeline, where intervention timing decisions can influence subsequent brushing times, allowing us to assess \emph{coverage} under evolving patterns of brushing activity.

\end{itemize}
The dataset was collected as part of a micro-randomized trial (MRT) \cite{klasnja2015microrandomized}, a study design in which intervention decisions are randomized repeatedly over time. In this trial, opportunities to deliver engagement prompts occur at predefined times throughout the day, and at each such time the system randomizes \textit{whether} to deliver a prompt. Intervention decisions were made twice per day, once in the morning (04:00) and once in the evening (16:00). At each decision time, the system randomized whether to deliver an engagement prompt. When a prompt was delivered, it was scheduled at user-provided brushing times, specified separately for morning and evening, and for weekdays and weekends.

A total of 79 participants were enrolled in the trial, with participants entering incrementally at a rate of approximately five participants every two weeks over the trial period from September 2023 to July 2024. Each participant contributed up to 70 days of data, with two intervention decision times per day, resulting in up to 140 decision times per participant.
Before the start of the trial, participants provided their usual brushing times separately for morning and evening, and for weekdays and weekends. These self-reported times were used directly as the intervention times, determining \emph{when} to deliver an intervention conditional on a decision to send a prompt at the corresponding decision time. These times were not updated and remained fixed throughout the trial. Participants used a commercial electronic toothbrush connected via Bluetooth to a mobile device. For analysis, brushing events were assigned to predefined time windows, with the morning window defined as 04:00–15:59 and the evening window defined as 16:00–03:59 the following day. Brushing data, including timestamps and duration measures, were recorded automatically and uploaded to a commercial secure server when the toothbrush was docked for charging. These data were processed and typically made available to the system prior to the next day’s decision window, but not in real time following brushing. Data from ten participants were excluded from the analysis due to synchronization issues between participant's devices and the study infrastructure, which resulted in intervention messages being delivered outside the intended 70-day intervention period. Table~\ref{tab:dataset_stats} summarizes descriptive statistics for the trial population and collected brushing data.

\begin{table}[t]
\centering

\begin{threeparttable}
\small
\setlength{\tabcolsep}{4pt}
\renewcommand{\arraystretch}{1.08}
\begin{tabular}{p{0.36\linewidth}
                >{\centering\arraybackslash}p{0.24\linewidth}
                >{\centering\arraybackslash}p{0.30\linewidth}}
\hline
\textbf{Property}
& \textbf{All Participants Combined}
& \textbf{Across Participants (mean $\pm$ SD)} \\
\hline
Number of participants & 69 & -- \\
Number of time windows & 9660 & 140 $\pm$ 0 \\
\% of time windows with brushing event & 56.4\% & 56.4 $\pm$ 23.5\% \\
\% of time windows with intervention prompt sent & 34.3\% & 34.3 $\pm$ 6.2\% \\

\hline
\end{tabular}

\end{threeparttable}
\caption{Descriptive statistics of the trial population and brushing data. Each day is divided into two 12-hour time windows: a morning window (04:00–15:59) and an evening window (16:00–03:59 of the following day). The ``All Participants Combined'' column reports aggregate values across all participants, while the ``Across Participants'' column reports per-participant statistics, computed over each participant’s trial period and summarized as mean $\pm$ standard deviation across participants.}
\label{tab:dataset_stats}
\end{table}

\section{Motivating Adaptive Intervention Timing }
\label{sec:motivation}

We analyze brushing-time events from the deployed trial described in the previous section in relation to the intervention timing used during deployment. This analysis motivates the need for adaptive intervention timing by showing that brushing times often differ substantially from user-specified times used to schedule interventions.

First, we quantify how often interventions were delivered after brushing had already occurred, indicating that the prompt arrived too late to be useful. Second, we introduce \emph{coverage} as a measure of intervention timing quality, capturing whether an intervention is delivered sufficiently close in time prior to a brushing event to be considered relevant. Together, these analyses illustrate multiple forms of timing misalignment under fixed intervention timing.

\subsection{Interventions Delivered After Brushing Events}

In the deployed trial, decisions on whether to deliver an intervention were made twice per day, once in the morning (04:00) and once in the evening (16:00), corresponding to predefined time windows (04:00–15:59 and 16:00–03:59 of the following day). At each decision time, the system randomized the decision of \emph{whether} to deliver an intervention prompt.
Across all participants, there were 9,660 time windows (69 participants $\times$ 140 windows each), of which 3,319 included an intervention prompt. Among these, 1,808 time windows contained both an intervention delivery and a brushing event.

Within these 1,808 windows, brushing occurred before the intervention in 47.6\% of cases when aggregated across all individuals. When summarized at the participant level, the proportion of such late interventions was 47.2 $\pm$ 21.8\%, indicating substantial variability across participants. This variability suggests that while some participant’s brushing times remained closely aligned with their reported usual times, others experienced frequent misalignment. Table~\ref{tab:timing_misalignment} summarizes these timing alignment statistics. Interventions delivered after brushing has already occurred are unlikely to influence that brushing event, as the opportunity to impact that event has already passed. These results highlight that, for a substantial subset of participants, interventions may have been ineffective not due to their content or delivery mechanism, but because they were delivered too late.

\begin{table}[t]
\centering

\small
\setlength{\tabcolsep}{4pt}
\renewcommand{\arraystretch}{1.08}
\begin{tabular}{p{0.54\linewidth}
                >{\centering\arraybackslash}p{0.16\linewidth}
                >{\centering\arraybackslash}p{0.24\linewidth}}
\hline
\textbf{Metric}
& \textbf{All Participants}
& \textbf{Across Participants (mean $\pm$ SD)} \\
\hline

Number of time windows with an intervention prompt (count)
& 3{,}319
& 48 $\pm$ 8.7 \\

Number of time windows with both intervention delivery and brushing event (count)
& 1{,}808
& 26.2 $\pm$ 11.8 \\

\% of time windows where brushing occurred before intervention (within the above 1,808 windows)
& 47.6\%
& 47.2 $\pm$ 21.8\% \\

Average delay between intervention and brushing (hours, conditional on brushing after intervention)
& 2.2
& 2.4 $\pm$ 1.4 \\

\hline
\end{tabular}
\caption{Intervention timing misalignment under fixed intervention times in the deployed trial. Rows report metrics with varying units, including counts (number of time windows), percentages, and time in hours. Values in the ``Across Participants'' column are reported as mean $\pm$ standard deviation across participants.}
\label{tab:timing_misalignment}
\end{table}

\subsection{Coverage-Based Evaluation of Intervention Timing}
\label{sec:coverage}

While late interventions capture one form of timing failure, they do not fully characterize whether an intervention is appropriately timed relative to brushing. In principle, one could eliminate late interventions by scheduling prompts sufficiently early in advance of expected brushing times (e.g., at the start of the morning or evening window). However, such interventions may occur too far in advance of brushing to meaningfully influence it. This highlights that avoiding late interventions alone is insufficient, and motivates the need for a more general way to quantify how well intervention times align with brushing events.

To capture this more generally, we introduce the notion of \emph{coverage}. Intuitively, coverage reflects whether an intervention occurs sufficiently close in time prior to a brushing event to be considered relevant. Throughout the paper, coverage is computed with respect to candidate intervention times, defined as the scheduled intervention times under a given timing strategy, regardless of whether a prompt was actually delivered.

Formally, let $I_{i,t}$ denote the candidate intervention time for individual $i$ at decision time $t$, and let $B_{i,t}$ denote the time of the first brushing event that occurs after $I_{i,t}$, if any. For a given tolerance parameter $C > 0$ (in hours), the intervention time is said to \emph{cover} a brushing event if
\[
0 < B_{i,t} - I_{i,t} \leq C.
\]

Smaller values of $C$ correspond to stricter timing requirements, while larger values allow greater temporal flexibility between the intervention time and the subsequent brushing event. Intervention times that occur after brushing (i.e., $B_{i,t} - I_{i,t} \leq 0$) are not counted as covered for any value of $C$. Coverage is computed by conditioning on time windows in which a brushing event is observed.  For each participant, we compute coverage as the proportion of such windows for which the corresponding candidate intervention time satisfies the coverage condition. Because the frequency of brushing varies substantially across participants (Table~\ref{tab:dataset_stats}), we summarize coverage at the individual level by reporting the mean and standard deviation across participants. Table~\ref{tab:coverage_motivation} reports coverage values for the fixed, user-provided intervention timing strategy used in the deployed trial across several values of $C$.  Even for a relatively permissive tolerance of $C = 5$ hours, the mean coverage across participants is 52.64\% ($\pm$ 20.97\%). That is, even with a tolerance of five hours within a 12-hour time window, only about half of such windows are covered on average. This indicates that fixed, user-provided intervention times frequently fail to align with actual brushing times, even under relatively lenient timing criteria. In addition, the consistently large standard deviation across all values of $C$ indicates substantial variability in coverage across participants. While some participants exhibit relatively high alignment between intervention times and brushing events, others experience consistently poor alignment. This heterogeneity suggests that fixed, user-provided intervention times may work reasonably well for some individuals but fail significantly for others.

\begin{table}[t]
\centering
\small
\begin{tabular}{lc}
\hline
\textbf{Tolerance parameter ($C$)}
& \textbf{Coverage (\%, mean $\pm$ SD)} \\
\hline
$C = 1$ hour  & 18.25 $\pm$ 12.88 \% \\
$C = 2$ hours & 32.75 $\pm$ 16.71 \% \\
$C = 3$ hours & 42.53 $\pm$ 20.34 \% \\
$C = 4$ hours & 48.50 $\pm$ 20.47 \% \\
$C = 5$ hours & 52.64 $\pm$ 20.97 \% \\
\hline
\end{tabular}
\caption{Coverage under fixed, user-provided intervention time strategy for different values of the tolerance parameter $C$. Coverage is computed with respect to candidate intervention times, using time windows in which brushing is observed. Values are reported as mean $\pm$ standard deviation across participants.}
\label{tab:coverage_motivation}
\end{table}

\begin{figure*}[h]
    \centering
    \begin{subfigure}[t]{0.49\textwidth}
        \centering
        \includegraphics[width=\linewidth]{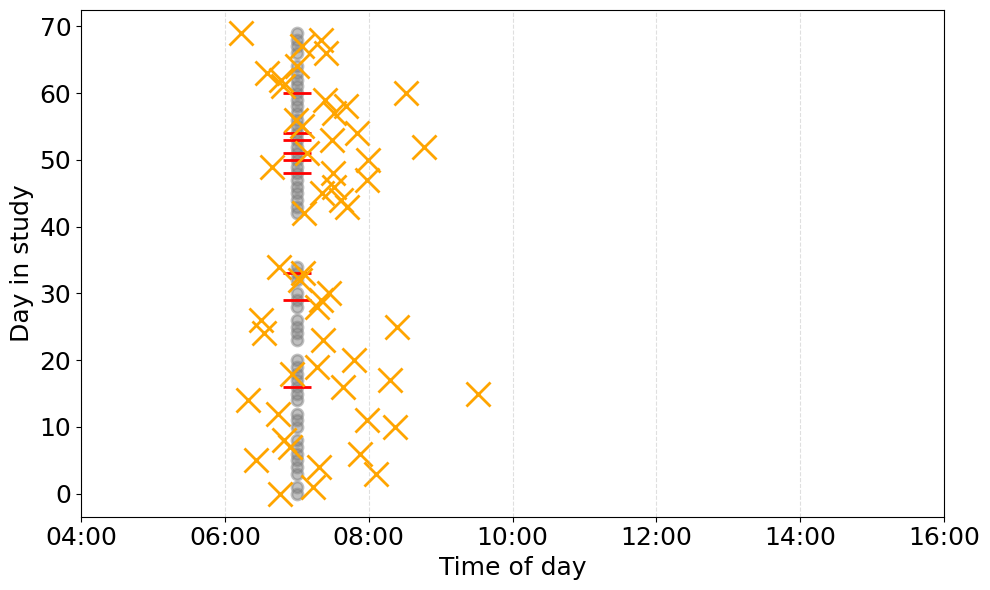}
        \caption{Participant A}
        \label{fig:participant_a}
    \end{subfigure}
    \hfill
    \begin{subfigure}[t]{0.49\textwidth}
        \centering
        \includegraphics[width=\linewidth]{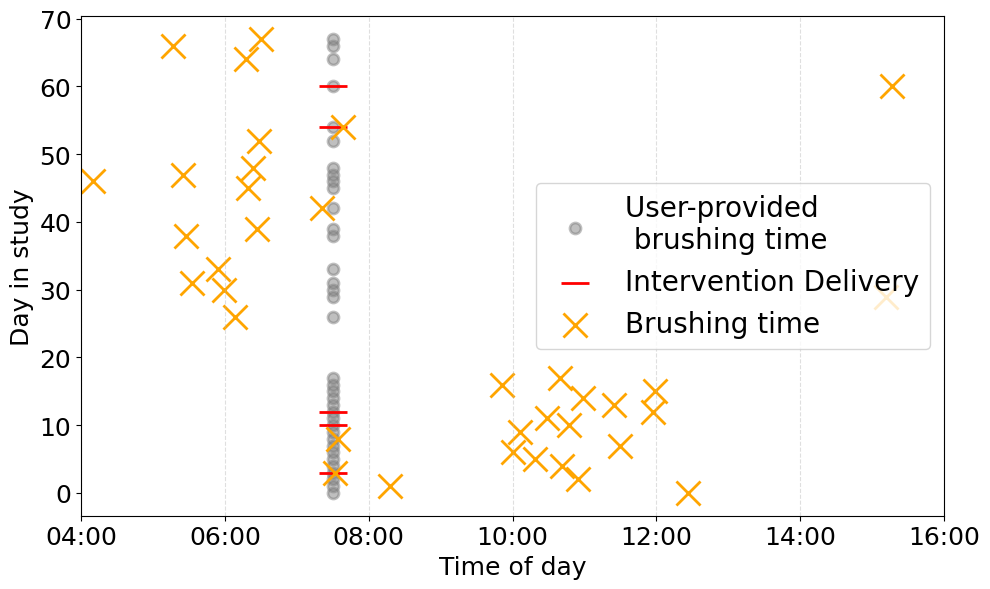}
        \caption{Participant B}
        \label{fig:participant_b}
    \end{subfigure}
    \hfill
    \begin{subfigure}[t]{0.49\textwidth}
        \centering
        \includegraphics[width=\linewidth]{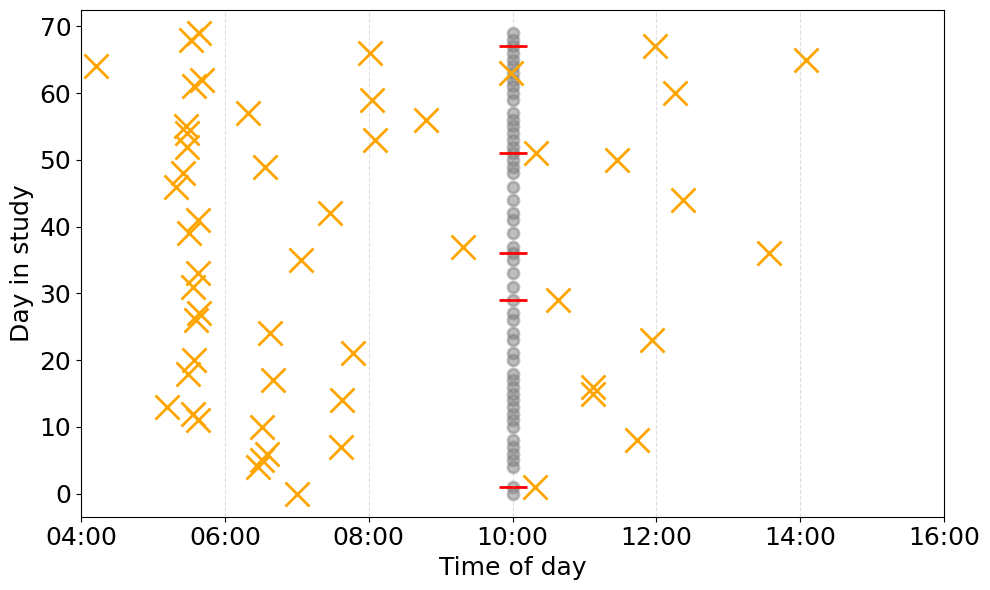}
        \caption{Participant C}
        \label{fig:participant_c}
    \end{subfigure}
        \hfill
    \begin{subfigure}[t]{0.49\textwidth}
        \centering
        \includegraphics[width=\linewidth]{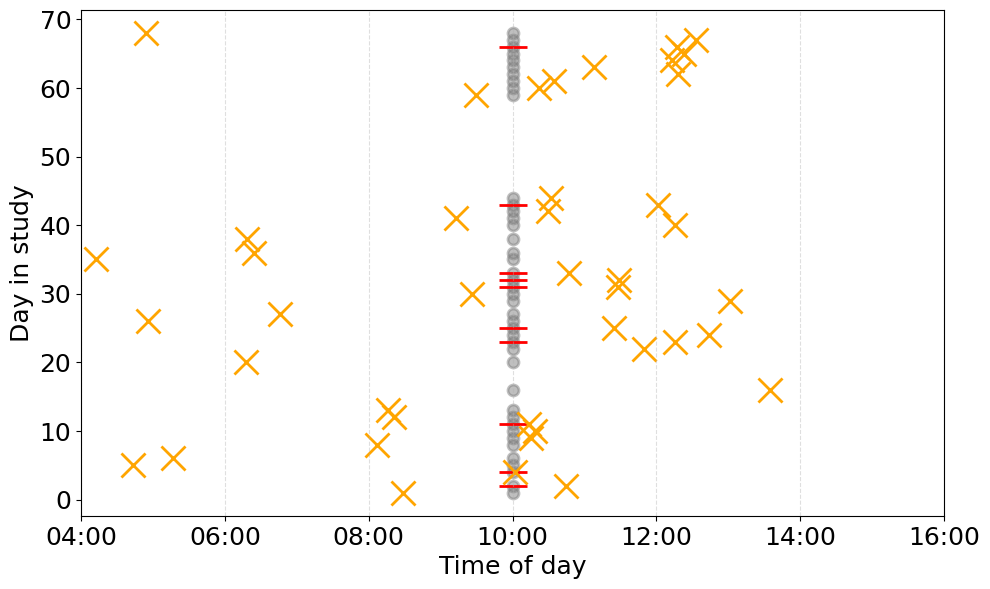}
        \caption{Participant D}
        \label{fig:participant_d}
    \end{subfigure}

\caption{Illustrative examples of participant-level brushing-time patterns for the morning time windows over the 70-day study period, shown relative to participant-provided usual brushing times used to set intervention timing.  (a) Participant whose observed brushing times remain close to their reported usual brushing time throughout the study, (b) participant whose brushing times exhibit gradual temporal drift over the course of the study, and (c) participant whose observed brushing times consistently deviate from their reported usual brushing times. (d) participant whose brushing times exhibit high variability, occurring both earlier and later than their reported usual brushing time across days.}

    \label{fig:participant_examples}
\end{figure*}

This variability is further illustrated in Figure~\ref{fig:participant_examples}, where we provide illustrative examples of participant-level brushing-time patterns over the 70-day study period for four participants from this dataset. The first participant exhibits brushing times that remain consistently close to their self-reported usual times, the second participant shows a gradual drift in brushing time over the course of the study, the third participant consistently brushes far from their reported times, typically much earlier than their stated usual brushing times, and the fourth participant exhibits high variability in brushing times, occurring both earlier and later than their reported usual time across days. Together, these examples highlight both the presence of timing misalignment and the diversity of brushing-time patterns across participants, reinforcing the need for adaptive intervention timing strategies.

\section{Adaptive Intervention Timing}
\label{Sec:adaptive_timing}

The analyses in the previous section highlight the limitations of fixed intervention timings based on user-provided input. Observed brushing times often vary substantially from reported habitual times, indicating that static schedules may become poorly aligned as routines evolve. To address this challenge, we propose two adaptive intervention timing strategies that update intervention times over the course of a study in response to observed brushing-time patterns.

In the remainder of this section, we describe these approaches and their implementation within an online decision-making framework. In the next section, we evaluate the proposed strategies using the dataset described in Section~\ref{Sec:dataset}. These evaluations inform the selection of an adaptive intervention timing approach for deployment in an ongoing clinical trial. More broadly, the evaluation framework introduced here is general and can be extended to other habitual behaviors to support assessment and refinement prior to real-world deployment.

\subsection{Modeling Habitual Behavior}

To enable adaptive intervention timing, we  require predictions of when an individual is likely to perform the target habitual behavior  (here, brushing as the target behavior in this case study). This necessitates individual-specific models that can generate behavior-timing predictions from the beginning of a study and update continuously as new behavioral data are observed. Importantly, these models must be trained and updated online, reflecting the sequential nature of the data available during real-world deployment.

This setting presents two key challenges. First, as reflected in the dataset described in Section~\ref{Sec:dataset}, each individual contributes a limited number of observations over the course of the intervention period, with at most 140 time windows per participant. Learning individual brushing-time patterns and their evolution over time from such sparse data is inherently difficult, particularly given the relatively short duration of typical intervention studies.

Second, brushing does not occur in every time window. When no brushing event is observed within a window, the exact brushing time is unobserved, resulting in right-censored observations that cannot be treated as fully observed outcomes. These censored observations must be appropriately accounted for during model training and updating.

Together, these characteristics place adaptive intervention timing in a sparse, online learning regime that requires models capable of incremental updating and robust uncertainty handling. In the following subsections, we describe the predictive models used to generate brushing-time predictions under these constraints and discuss how their outputs are incorporated into the proposed adaptive intervention timing frameworks.

\subsection{Intervention Timing Strategies}

In the settings considered in this work, intervention decisions are made \emph{ahead of time}, prior to the occurrence of the target behavior. For each time window \(t\), the system must determine (i) whether to deliver an intervention and (ii) if so, when the intervention should be delivered. While both are components of the overall online decision-making process, this paper focuses specifically on the problem of \emph{intervention timing}.

Intervention timing strategies determine when to schedule prompts relative to predicted times at which the target behavior is likely to occur. In this work, we focus on approaches that leverage estimates of expected behavior time and their associated uncertainty to set intervention times. Even when an intervention is not delivered, each strategy defines a corresponding candidate intervention time, representing when an intervention would have been scheduled under that strategy.

We consider four intervention timing strategies. The first corresponds to the baseline strategy used in the deployed trial, which is non-adaptive and based solely on participant-provided usual times. In addition, we evaluate three alternative strategies: one non-adaptive strategy that applies a fixed offset to user-provided times, and two adaptive strategies that rely on online predictive models. These adaptive strategies use, for each time window \(t\), an estimate of the expected behavior time \(\hat{\mu}_t\) and its associated uncertainty \(\hat{\sigma}_t\), which are updated over time as new data are observed. We describe these strategies in detail below.

\paragraph{User Input (Baseline) .}
This strategy corresponds to the intervention timing used in the deployed trial (Section~\ref{Sec:dataset}). Intervention times are set directly to the participant-reported habitual times and remain fixed throughout the study. That is, the intervention time is given by
\begin{equation}
I_t = U_t,
\end{equation}
where \(U_t\) denotes the participant-reported habitual time for time window  \(t\). This baseline reflects a non-adaptive, user-specified timing strategy.

\paragraph{User Input with Fixed Offset (User+ Offset).}
This strategy extends the baseline by introducing a fixed temporal offset to encourage interventions to precede the expected behavior. Specifically, the intervention time is scheduled earlier than the reported habitual time:
\begin{equation}
I_t = U_t - \Delta,
\label{eq:user_offset}
\end{equation}
where \(\Delta\) is a fixed offset that does not adapt over time. In our evaluation, \(\Delta\) is treated as a hyperparameter; details of its selection are described in Section~\ref{sec:resultsandeval}.

\paragraph{Model Mean with Fixed Offset (Mean + Offset).}
This strategy uses model-based predictions of the expected behavior time and applies a fixed temporal offset. For each time window \(t\), the predictive model provides an estimate of the expected behavior time \(\hat{\mu}_t\). The intervention time is then set as
\begin{equation}
I_t = \hat{\mu}_t - \Delta,
\label{eq:model_fixed}
\end{equation}
where \(\Delta\) is a fixed offset. Unlike the previous strategy, this approach allows intervention times to adapt to changes in the predicted behavior time over the course of the study. In our evaluation, \(\Delta\) is treated as a hyperparameter; details of its selection are described in Section~\ref{sec:resultsandeval}.

\paragraph{Model Mean with Uncertainty (Mean + Unc).}
When using model-based predictions to set intervention times, we operate in a low-data setting where estimates of behavior time may be highly uncertain. To balance the goal of delivering interventions before the behavior while remaining as close as possible to the expected behavior time, this strategy incorporates both the predictive mean and its associated uncertainty. Intuitively, when the model is more certain, interventions are scheduled closer to the predicted time, whereas higher uncertainty leads to earlier interventions to reduce the risk that the prompt occurs after the behavior. Following this rationale \cite{gazi2025sigmascheduling}, the intervention time is set as
\begin{equation}
I_t = \hat{\mu}_t - \alpha \hat{\sigma}_t,
\label{eq:uncertainty_timing}
\end{equation}
where \(\alpha\) is a hyperparameter that controls how conservatively the intervention is scheduled relative to predictive uncertainty. Details of its selection are provided in Section~\ref{sec:resultsandeval}.

\subsection{Online Predictive Model Choices}
\label{sec:onlinepredmodel}
Building on the discussion above, adaptive intervention timing requires predictive models that can operate in a low-data, online setting and provide both an estimate of behavior time and an associated measure of uncertainty. In our setting, each individual contributes a limited number of observations over time, and the target behavior may be absent in many time windows due to censoring. To address this, we model the time of the behavior conditional on its occurrence within a predefined time window. That is, for windows in which the behavior is observed, we use the observed time to train predictive models. This modeling choice is aligned with the intervention timing objective considered in this work: determining when to schedule an intervention relative to when the behavior is expected to occur. By conditioning on windows with observed behavior, we focus directly on estimating behavior time, while avoiding the need to jointly model both occurrence and timing in a sparse, censored data setting.

As a result, predictive models must be able to update incrementally as new data arrive, remain stable under sparse observations, and, when possible, benefit from information learned across individuals. In addition, because intervention timing decisions depend not only on expected behavior time but also on the risk of mistiming, these models must provide both an estimate of behavior time and a corresponding measure of uncertainty.

To this end, we consider three classes of online predictive models that differ in complexity, inductive bias, and how uncertainty is quantified: online Bayesian linear regression, an online decision tree, and an online neural network with Monte Carlo dropout.

\paragraph{Online Bayesian Linear Regression (BLR).}
The online BLR model provides a probabilistic framework well suited to low-data settings. For each individual, we model behavior time as a linear function of contextual features with Gaussian noise. As new observations are collected, posterior distributions over model parameters are updated sequentially using closed-form Bayesian updates. The estimate of behavior time is given by the posterior mean, and uncertainty is derived from the posterior predictive variance, which reflects both observation noise and parameter uncertainty. This explicit uncertainty quantification makes BLR a strong baseline for adaptive intervention timing.

\paragraph{Online Decision Tree (DT).}
We also consider an online decision tree model based on a Hoeffding Tree regressor, which incrementally updates its structure as new data arrive. Each individual’s data are processed sequentially, and splits are introduced only when there is sufficient statistical evidence to support them. Predictions are generated from the leaf node reached by the current input. The estimate of behavior time is given by the leaf-level mean, and uncertainty is quantified using the variance of observed outcomes within that leaf, maintained via an online variance estimator. This approach provides a non-linear, interpretable alternative to BLR while retaining online learning capability.

\paragraph{Online Neural Network (NN) with Monte Carlo Dropout.}
Finally, we evaluate an online neural network model that uses Monte Carlo (MC) dropout to approximate predictive uncertainty \cite{gal2016dropout}. The network is trained incrementally using individual-specific data, and dropout is applied at inference time to generate multiple stochastic forward passes. The estimate of behavior time is computed as the average of these forward passes, while uncertainty is estimated from their empirical variance. Although neural networks typically require more data to perform well, MC dropout provides a practical mechanism for uncertainty estimation in an online setting and allows us to assess whether increased model expressiveness yields benefits for adaptive intervention timing under sparse data conditions.

Additional implementation details for all predictive models are provided in the Supplementary Material. We evaluate adaptive intervention timing based on predictions from these models against non-adaptive strategies in two complementary ways, as described in the following section.

\section{Evaluation and Results}
\label{sec:resultsandeval}

Having described the adaptive intervention timing strategies and the predictive models used to support them, we now turn to their evaluation. We use the dataset described in Section~\ref{Sec:dataset} to conduct two complementary types of evaluation.

In the first form of evaluation, we examine how different intervention timing strategies align with observed brushing times. Using data from the deployed study, we compare strategies based on how well they \emph{cover} observed brushing events.

Second, we conduct a simulation-based evaluation in which brushing times are generated rather than treated as fixed observations. Specifically, predictive models are fit to the deployed dataset and used as generative models to simulate brushing-time trajectories. This allows intervention timing decisions to influence subsequent brushing times, enabling evaluation of strategies in a setting where brushing times evolve in response to interventions. We evaluate performance using the same coverage-based metric, enabling comparison with the observational analysis.

Based on these evaluations, we selected an intervention timing strategy for deployment in an ongoing clinical trial. We conclude this section by presenting preliminary results from this deployment.

\subsection{Evaluation with Observed Brushing Times}
\label{sec:coveragetiming}

\begin{table*}[t!]
\centering

\small
\setlength{\tabcolsep}{6pt}
\renewcommand{\arraystretch}{0.95}

\resizebox{0.82\textwidth}{!}{
\begin{tabular}{c cc}
\toprule
\makecell[c]{\textbf{Tolerance} \\ \textbf{(\(C\))}}
& \multicolumn{2}{c}{\textbf{Intervention Timing Strategy}} \\
\cmidrule(lr){2-3}
& \makecell[c]{\textbf{Baseline} \\ \textbf{(deployed intervention timing= $U_{t}$)}}
& \makecell[c]{\textbf{User+ Offset} \\ $ (U_t - \Delta )$} \\
\midrule
\textbf{1 Hour} & 18.25 $\pm$ 12.88 & 18.25\% $\pm$ 12.88 \\
\textbf{2 Hour} & 32.75 $\pm$ 16.71\% & 34.23 $\pm$ 18.58 \\
\textbf{3 Hour} & 42.53 $\pm$ 20.34\% & $46.56 \pm 20.47 (+4.02)^{*}$ \\
\textbf{4 Hour} & 48.50 $\pm$ 20.47\% & $56.32 \pm 21.81 (7.81)^{*}$ \\
\textbf{5 Hour} & 52.64 $\pm$ 20.97\% & $63.39 \pm 21.13 (+10.74)^{*}$ \\
\bottomrule
\end{tabular}
}
\captionof{table}{Coverage (\%) comparison for two non-adaptive intervention timing strategies across different tolerance values \(C\). The \emph{Baseline (User Input)} strategy corresponds to the deployed intervention timing, where the intervention time is set equal to the participant-reported brushing time, while the \emph{User + Offset} strategy shifts these times earlier by a fixed offset \(\Delta\). Values are reported as mean \(\pm\) standard deviation across participants. Parentheses indicate the mean paired difference relative to the baseline for the same value of \(C\). \(^{*}\) indicates a statistically significant difference compared to the baseline based on a paired Wilcoxon signed-rank test (\(p < 0.05\)).}

\label{tab:coverage_baseline_userinput}

\vspace{0.8cm}

\small
\setlength{\tabcolsep}{5pt}
\renewcommand{\arraystretch}{0.95}

\resizebox{\textwidth}{!}{
\begin{tabular}{c l cc}
\toprule
\makecell[c]{\textbf{Tolerance} \\ \textbf{(\(C\))}}
& \textbf{Predictive Model Choice}
& \multicolumn{2}{c}{\textbf{ Intervention Timing Strategy}} \\
\cmidrule(lr){3-4}
& &
\makecell[c]{\textbf{Mean+ Offset} \\ {($ \hat{\mu}_t - \Delta $)} }
& \makecell[c]{\textbf{Mean+ Unc} \\ ( $\hat{\mu}_t - \alpha \hat{\sigma}_t$) }\\
\midrule

\multirow{3}{*}{\textbf{1 Hour}}
& Online Bayesian Linear Regression (BLR) &  \textcolor{red}{$21.69 \pm 12.63 (+3.44)^{*}$} &  \textcolor{red} {$21.36 \pm 12.76 (+3.10)^{*}$} \\
& Online Decision Tree (DT)  & $18.22 \pm 9.48$  & $18.53 \pm 10.65$ \\
& Online Neural Network (NN)  & $18.43 \pm 10.48$ & $18.17 \pm 10.17$ \\
\midrule

\multirow{3}{*}{\textbf{2 Hour}}
& Online BLR & \textcolor{red}{$39.21 \pm 18.42 (+6.46)^{*}$} & \textcolor{red}{$40.37 \pm 17.29 (+7.62)^{*}$} \\
& Online DT  & $34.11 \pm 15.14$ & $34.44 \pm 14.83$ \\
& Online NN  & $35.20 \pm 14.70$ & $35.92 \pm 16.37 (+3.17)^{*}$ \\
\midrule

\multirow{3}{*}{\textbf{3 Hour}}
& \textcolor{red}{Online BLR} & \textcolor{red}{$53.76 \pm 19.50\;(+11.22)^{*}$} & \textcolor{red}{$52.42 \pm 18.84\;(+9.88)^{*}$} \\
& Online DT  & $47.90 \pm 16.53 (+5.37)^{*} $ & $47.12 \pm 18.93 (+4.59)^{*} $ \\
& Online NN  & $45.64  \pm 16.50 (+3.11)^{*} $ & $48.46 \pm 18.97 (+5.93)^{*} $ \\
\midrule

\multirow{3}{*}{\textbf{4 Hour}}
& \textcolor{red}{Online BLR} & \textcolor{red}{$64.19 \pm 18.93\;(+15.64)^{*}$} & \textcolor{red}{$63.30 \pm 17.23\;(+14.79)^{*}$} \\
& Online DT  & $58.60 \pm 18.42\;(+10.10)^{*}$ & $57.67 \pm 18.44\;(+9.17)^{*}$ \\
& Online NN  & $60.15 \pm 17.95\;(11.65)^{*}$ & $59.97 \pm 18.37\;(+11.47)^{*}$ \\
\midrule

\multirow{3}{*}{\textbf{5 Hour}}
& \textcolor{red}{Online BLR} & \textcolor{red}{$71.20 \pm 17.06\;(+18.56)^{*}$} & \textcolor{red}{$70.42 \pm 15.43\;(+17.77)^{*}$} \\
& Online DT  & $66.35 \pm 17.09\;(+13.71)^{*}$ & $66.47 \pm 17.25\;(+13.83)^{*}$ \\
& Online NN  & $65.72 \pm 18.72\;(+13.08)^{*}$ & $66.34 \pm 17.25\;(+13.70)^{*}$ \\
\bottomrule
\end{tabular}
}
\captionof{table}{Coverage (\%) comparison for adaptive intervention timing strategies across predictive model choices and tolerance values \(C\). The two adaptive strategies correspond to \emph{Model Mean with Fixed Offset} (Mean + Offset) and \emph{Model Mean with Uncertainty} (Mean + Unc.). Values are reported as mean \(\pm\) standard deviation across participants. Parentheses indicate the mean paired difference relative to the deployed baseline for the same value of \(C\) and intervention timing strategy. \(^{*}\) indicates a statistically significant difference compared to the baseline based on a paired Wilcoxon signed-rank test (\(p < 0.05\)). Results highlighted in \textcolor{red}{red} denote the best-performing model and intervention timing strategy combination for each tolerance value among those that are statistically significant over the baseline.}
\label{tab:coverage_model_based}
\end{table*}
We first evaluate the intervention timing strategies using observed brushing times from the deployed trial. In this evaluation, brushing times are treated as fixed observations and are not influenced by the intervention timing strategy. This allows us to assess how well different strategies align with observed brushing times under the assumption that brushing patterns remain unchanged. For each intervention timing strategy, we consider time windows in which a brushing event is observed. Within each such window, we compute the candidate intervention time and determine whether it covers the corresponding brushing event according to the definition of coverage. Coverage is computed at the individual level and then summarized across participants. The results are reported in Tables~\ref{tab:coverage_baseline_userinput} and~\ref{tab:coverage_model_based}.

For participant-specific predictive models, we adopt a leave-one-subject-out (LOSO) approach to ensure fair evaluation and prevent information leakage. Specifically, for each participants in the trial, model parameters and priors are estimated using data from all other participants. Similarly, hyperparameters associated with intervention timing strategies, such as the fixed offset \(\Delta\) and the uncertainty scaling parameter \(\alpha\), are selected on a per-participant basis using the same LOSO procedure. These hyperparameters are chosen to optimize the evaluation metric (coverage) based on data from other participants. We evaluated an extensive set of statistical and hand-crafted features for the predictive models prior to feature selection. The final selected feature set used for the online BLR and DT models is summarized in Table~\ref{tab:features}. Additional details on feature construction and model specifications, including those for the NN, are provided in the appendix.
\begin{table}[t]
\centering
\small
\setlength{\tabcolsep}{4pt}
\renewcommand{\arraystretch}{1.08}
\begin{tabular}{p{0.28\linewidth} p{0.66\linewidth}}
\hline
\textbf{Feature} & \textbf{Description} \\
\hline
Time of day & Indicator for morning or evening time window \\
Day of week & Indicator for weekday or weekend \\
Recent brushing time (same window) & Most recent observed brushing time in the same time window \\
Earliest brushing time (past 7 days, same window) & Earliest observed brushing time in the same time window over the past 7 days \\
Latest brushing time (past 7 days, same window) & Latest observed brushing time in the same time window over the past 7 days \\
Exponential average (past 7 days, same window) & Exponentially weighted average of brushing times in the same time window over the past 7 days \\
Coefficient of variation (past 7 days, same window) & Ratio of standard deviation to mean of brushing times in the same time window over the past 7 days \\
Zero brushing count (past 7 days, same window) & Number of time windows in the past 7 days (same window) in which brushing did not occur \\
User-provided time & Participant-reported brushing time for the current time window (morning/evening, weekday/weekend) \\
\hline
\end{tabular}
\caption{Final selected feature set used for online BLR and DT models after evaluating a broader set of statistical and hand-crafted features. Details of the full feature set are provided in the appendix.}
\label{tab:features}
\end{table}

Table~\ref{tab:coverage_baseline_userinput} compares the coverage of the deployed intervention timing (\emph{Baseline (User Input)}) with a non-adaptive strategy that applies a fixed offset (\emph{User + Offset}). For the smallest tolerance value (\(C = 1\)), the optimal offset is \(\Delta = 0.00 \pm 0.00\) across participants, resulting in identical coverage between the two strategies. Starting from \(C = 2\), we observe improvements in coverage when applying a fixed offset, although these gains are modest (\(\Delta = 0.50 \pm 0.02\)) and not statistically significant. For larger tolerance values (\(C \geq 3\)), the gains become more pronounced and statistically significant. At \(C = 3\), the fixed-offset strategy yields an improvement of \(+4.02\%\) over the baseline, with a mean offset of \(\Delta = 1.00 \pm 0.05\) across participants. This trend continues for \(C = 4\) and \(C = 5\), with mean offsets of \(\Delta = 1.00 \pm 0.03\) and \(\Delta = 2.00 \pm 0.04\), resulting in coverage improvements of \(+7.81\%\) and \(+10.74\%\), respectively. These results indicate that shifting intervention times earlier than user-provided schedules can improve alignment with brushing times, particularly as greater flexibility is allowed through larger tolerance values.

In Table~\ref{tab:coverage_model_based}, we evaluate adaptive intervention timing strategies across different predictive models for estimating brushing time, where models are updated online at the individual level. We make the following observations:

\begin{itemize}

    \item Results highlighted in red denote, for each tolerance value, the model and intervention timing strategy combination that achieves the highest coverage among those that are statistically significant relative to the deployed baseline intervention timing. Across all tolerance values, these best-performing results consistently correspond to the Online Bayesian Linear Regression (BLR) model, which achieves higher coverage than the Online Decision Tree (DT) and Online Neural Network (NN) models regardless of the intervention timing strategy. This suggests that simpler models may be more effective for estimating brushing time in low-data, personalized settings.

    \item For each tolerance value, adaptive intervention timing strategies outperform baseline intervention timing, with the performance gap increasing as \(C\) increases. For example, under the BLR model, coverage improves from \(42.53\%\) ( \(C = 3\)) to \(53.76\%\) and \(52.42\%\) for Mean + Offset and Mean + Unc., respectively. This improvement becomes more pronounced at higher tolerance values, reaching \(71.20\%\) and \(70.42\%\) at \(C = 5\), corresponding to gains exceeding \(18\%\) over the baseline.

      \item Across both adaptive intervention timing strategies—one based on a fixed offset \(\Delta\) from the predicted brushing time and the other incorporating predictive uncertainty—we observe no statistically significant differences in the coverage metric defined in this paper.

\end{itemize}

Overall, these results highlight the importance of adapting intervention timing to individual brushing-time patterns. While simple non-adaptive adjustments, such as fixed offsets, can partially mitigate timing misalignment, adaptive strategies that leverage predictive models provide more consistent improvements in coverage. These gains become more pronounced as the tolerance \(C\) increases, where greater flexibility in intervention timing allows better alignment with brushing times. Among the models considered, the Online BLR approach provides the most reliable performance gains, suggesting that simpler models may be better suited for estimating brushing time in low-data, personalized settings. These findings motivate the need to evaluate adaptive intervention timing in settings where brushing times are not fixed and may evolve in response to interventions, which we examine next through simulation-based analysis.

\subsection{Evaluation with Simulated Brushing Times}
\label{sec:simulation}
In the previous evaluation, we assessed intervention timing strategies based on coverage with respect to observed brushing times from the deployed trial. In contrast, we now consider a setting in which intervention timing decisions can influence subsequent brushing times. Specifically, we evaluate intervention timing strategies using the same coverage-based metric in a simulated environment, where brushing times evolve over time in response to intervention timing decisions.

To conduct this evaluation, we adopt a simulation-based approach. For each participant in the deployed trial, we construct a generative model by fitting predictive models to that participant’s historical data. These fitted models are then used to simulate brushing-time trajectories and other relevant signals over time. At a high level, we simulate participant trajectories sequentially, including brushing times and app engagement. App engagement is included as it serves as an input to the participant-specific models used to generate brushing times in the simulation. For each time window  \(t\), a candidate intervention time is determined by the intervention timing strategy. The system then evolves sequentially, using simulated outcomes to update the participant’s history, which in turn informs future predictions. The decision of whether to deliver an intervention is governed by a fixed stochastic policy, where an intervention is delivered with probability \(0.5\) for each time window. The rationale for this choice is discussed in the subsequent subsection. Performance is evaluated using the same coverage-based metric defined earlier. Here, coverage is computed with respect to the candidate intervention times and the corresponding simulated brushing times.
\begin{table*}[t]
\centering
\small
\setlength{\tabcolsep}{4pt}
\renewcommand{\arraystretch}{0.95}

\resizebox{\textwidth}{!}{
\begin{tabular}{c c cccc}
\toprule
\makecell[c]{\textbf{Tolerance} \\ \textbf{(\(C\))}}
& \textbf{Simulation Variant}
& \multicolumn{4}{c}{\textbf{Intervention Timing Strategy}} \\
\cmidrule(lr){3-6}
&
& \makecell[c]{Baseline  \\ ($U_{t}$)}
& \makecell[c]{User + Offset \\ ($U_{t}-\Delta$)}
& \makecell[c]{Mean + Offset \\ ($\hat{\mu}_t - \Delta $)}
& \makecell[c]{Mean + Unc \\ ($\hat{\mu}_t - \alpha \hat{\sigma}_t$)} \\
\midrule

\multirow{5}{*}{}
\textbf{1 Hour} &  & 14.10 $\pm$ 7.21\% & 14.92 $\pm$ 7.50 & 15.20 $\pm$ 6.92 & 14.96 $\pm$ 6.67 \\
\textbf{2 Hour} &  & 26.52 $\pm$ 13.04\% & 26.88 $\pm$ 13.45 & \textcolor{red} {30.43 $\pm$ 13.32 (+3.91)$^{*}$ } & \textcolor{red} { 29.54 $\pm$ 12.91 (+3.02)$^{*}$ } \\
\textbf{3 Hour} &  \textbf{Data-driven} & 37.60 $\pm$ 16.41\% & 38.99 $\pm$ 15.78 & \textcolor{red}{43.70 $\pm$ 17.43 (+6.10)$^{*}$} & \textcolor{red}{44.41 $\pm$ 16.42\% (+6.81)$^{*}$} \\
\textbf{4 Hour} &  & 46.06 $\pm$ 17.67\% & 49.64 $\pm$ 17.84 & \textcolor{red}{54.75 $\pm$ 18.79 (+8.69)$^{*}$} & \textcolor{red}{55.90 $\pm$ 18.66\% (+9.84)$^{*}$} \\
\textbf{5 Hour} &  & 51.43 $\pm$ 18.24\% & 58.67 $\pm$ 18.39 (+7.21)$^{*}$ & \textcolor{red}{63.79 $\pm$ 19.02 (+12.36)$^{*}$} & \textcolor{red}{65.69 $\pm$ 19.12\% (+14.26)$^{*}$} \\
\midrule

\multirow{5}{*}{}
\textbf{1 Hour}&  & 25.30 $\pm$ 10.01\% & 25.32 $\pm$ 9.51  & 25.98 $\pm$ 8.62  & 25.82 $\pm$ 8.80  \\
\textbf{2 Hour} &  & 37.11 $\pm$ 12.53\% & 37.17 $\pm$ 12.68 & \textcolor{red}{40.14 $\pm$ 11.58 (+3.03)$^{*}$} & \textcolor{red}{ 41.21 $\pm$ 11.33 (+4.10)$^{*}$} \\
\textbf{3 Hour} & \makecell[c]{\textbf{Moderate-effect} } & 46.31 $\pm$ 14.96\% & 48.88 $\pm$ 13.94 & \textcolor{red}{51.68 $\pm$ 13.96 (+5.37)$^{*}$} & \textcolor{red}{51.94 $\pm$ 13.85 (+5.63)$^{*}$} \\
\textbf{4 Hour} &  & 52.96 $\pm$ 16.28\% & 57.86 $\pm$ 14.52 (+4.90)$^{*}$ & \textcolor{red}{61.29 $\pm$ 15.34 (+8.33)$^{*}$} & \textcolor{red}{61.86 $\pm$ 15.78 (+8.90)$^{*}$} \\
\textbf{5 Hour} &  & 56.67 $\pm$ 16.29\% & 64.18 $\pm$ 14.64 (+7.51)$^{*}$ & \textcolor{red}{69.90 $\pm$ 14.50 (+13.23)$^{*}$ } & \textcolor{red}{69.45 $\pm$ 14.32 (+12.78)$^{*}$} \\
\midrule

\multirow{5}{*}{}
\textbf{1 Hour} &  & 28.10 $\pm$ 11.21\% & 27.71 $\pm$ 10.46 & 29.43 $\pm$ 10.67 & 29.33 $\pm$ 10.89 \\
\textbf{2 Hour} &  & 42.58 $\pm$ 13.91\% & 42.65 $\pm$ 13.90 & \textcolor{red}{47.54 $\pm$ 13.46  (+4.96)$^{*}$} & \textcolor{red}{48.32 $\pm$ 13.22 (+5.74)$^{*}$)}\\
\textbf{3 Hour} & \makecell[c]{\textbf{Strong-effect} } & 51.92 $\pm$ 16.17\% & 54.92 $\pm$ 14.40 (+3)$^{*}$ & \textcolor{red}{60.65 $\pm$ 14.43 (+8.73)$^{*}$ } & \textcolor{red}{60.43 $\pm$ 13.91 (+8.51)$^{*}$ }\\
\textbf{4 Hour} &  & 58.65 $\pm$ 16.29\% & 64.96 $\pm$ 14.67 (+6.31) $^{*}$ & \textcolor{red}{70.66 $\pm$ 14.61 (+12.01)$^{*}$} & \textcolor{red}{71.23 $\pm$ 14.23 (+12.58)$^{*}$} \\
\textbf{5 Hour} &  & 62.42 $\pm$ 16.76\% & 69.92 $\pm$ 14.30 (+7.50)$^{*}$ & \textcolor{red}{77.40 $\pm$ 12.91 (+14.98)$^{*}$} & \textcolor{red}{78.04 $\pm$ 13.04 (+15.62)$^{*}$} \\
\bottomrule
\end{tabular}
}

\caption{Simulation results of coverage across three variants that differ in how intervention timing influences subsequent brushing behavior. Coverage values are reported as mean \(\pm\) standard deviation across participants. Parentheses indicate the mean paired difference relative to the baseline for the same value of \(C\), and \(^{*}\) indicates statistical significance based on a paired Wilcoxon signed-rank test (\(p < 0.05\)). For adaptive intervention timing strategies, results are reported using the Online Bayesian Linear Regression (BLR) predictive model. Results highlighted in \textcolor{red}{red} denote, for each variant and tolerance value, the intervention timing strategy that achieves the highest coverage among those that are statistically significant compared to the baseline.}
\label{tab:simulation_combined}
\end{table*}

The simulation environment models two key components: (i) brushing time and (ii) participant app engagement. Brushing time is generated using participant-specific survival models, while app engagement is incorporated as an input feature to these models. Full details of these outcome-generating models are provided in the appendix.

To assess how adaptive intervention timing performs under different assumptions about the effect of intervention timing on brushing times, we consider three simulation variants. These variants differ in how intervention timing modifies the underlying brushing-time distribution beyond what is captured by the fitted survival models.

\begin{itemize}

    \item \textbf{Data-driven :} Brushing times are generated directly from the fitted survival model without any additional modification based on intervention timing. In this case, the influence of intervention delivery on brushing times is limited to what has been learned from the observed data.

    \item \textbf{Moderate-effect variant} When an intervention is delivered, the brushing-time distribution is modified to increase the probability of brushing within the \(C\)-hour window following the intervention time. This is achieved by locally adjusting the fitted survival distribution in this window, resulting in a moderate shift of brushing times toward the intervention time.

    \item \textbf{Strong-effect variant} When an intervention is delivered, the brushing-time distribution is modified more aggressively to substantially increase the probability of brushing within the \(C\)-hour window following the intervention time. This results in a stronger shift of brushing times toward the intervention time compared to Moderate-effect Variant.

\end{itemize}
These variants allow us to evaluate intervention timing strategies under different assumptions about the strength of the causal effect of intervention timing on brushing times.
For each simulation variant, we generate brushing-time trajectories sequentially for each participant. For each time window \(t\), a candidate intervention time is determined using the intervention timing strategy, and brushing time is simulated accordingly. Coverage is computed for each time window based on the candidate intervention time and the simulated brushing time, then aggregated at the individual level and summarized across participants. Results are reported in Table ~\ref{tab:simulation_combined}. To account for variability due to stochastic simulation, we repeat the simulation procedure multiple times. Specifically, for each simulation variant, we perform 100 independent simulation runs. For each participant, coverage is computed within each run and then averaged across the 100 runs. These participant-level averages are then aggregated across participants to obtain the reported mean and standard deviation.

Across all three simulation variants, we consistently observe that adaptive intervention timing strategies outperform the baseline as the tolerance parameter \(C\) increases. For stricter tolerance values (\(C = 1\)), improvements are minimal across all variants, indicating limited opportunity for timing adjustments to improve alignment under tight constraints. However, starting from \(C \geq 2\), adaptive strategies begin to yield statistically significant gains.

Under the data-driven variant, coverage improves from \(26.52\%\) to \(30.43\%\) and \(29.54\%\) at \(C = 2\), and further to \(63.79\%\) and \(65.69\%\) at \(C = 5\), corresponding to gains of up to \(14.26\%\) over the baseline. Similar trends are observed under the moderate-effect variant, where coverage increases from \(37.11\%\) to \(40.14\%\) and \(41.21\%\) at \(C = 2\), and up to \(69.90\%\) and \(69.45\%\) at \(C = 5\), yielding improvements exceeding \(13\%\). Under the strong-effect variant, the gains are even more pronounced, with coverage improving from \(42.58\%\) to \(47.54\%\) and \(48.32\%\) at \(C = 2\), and reaching \(77.40\%\) and \(78.04\%\) at \(C = 5\), corresponding to improvements of up to \(15.62\%\).

In addition, overall coverage increases across all strategies as the influence of intervention timing increases, moving from the data-driven variant to the moderate-effect and strong-effect variants. This trend is expected given the simulation design, where stronger modifications to the brushing-time distribution increase the likelihood of brushing within the tolerance window. Across all variants, the magnitude of improvement from adaptive strategies also increases with \(C\), indicating that adaptive timing is particularly beneficial when greater flexibility in intervention timing is allowed. Taken together, these results demonstrate that adaptive intervention timing improves alignment with brushing times under both fixed and intervention-influenced settings, with larger gains in environments where intervention timing more strongly shifts brushing times.

\subsection{Preliminary Results from a Randomized Controlled Trial of an Oral Health Digital Intervention}

Based on the results of the evaluations in the previous subsections, we deployed an adaptive intervention timing strategy to determine \textit{when} to intervene in an ongoing randomized controlled trial (RCT)\footnote{ClinicalTrials.gov identifier: NCT07167875, \url{https://clinicaltrials.gov/study/NCT07167875?viewType=Card&term=Oralytics&rank=2}.}. In this study, participants in the treatment group are scheduled to receive, on average, one intervention per day, with intervention decisions made twice daily—once in the morning (04:00) and once in the evening (16:00). For each time window, the system determines whether to deliver an engagement prompt with probability \(0.5\), ensuring an average of one intervention per day. This stochastic delivery mechanism is consistent with the setup used in our simulation-based evaluation, enabling a direct comparison between simulated and real-world deployment settings. When a prompt is delivered, its timing is determined using the adaptive intervention timing strategy.

The intervention content and delivery setup are consistent with those described in Section~\ref{Sec:dataset}. Based on the results from previous evaluations, we selected the uncertainty-aware intervention timing strategy combined with an Online BLR model for predicting brushing time. Although we did not observe a statistically significant difference between the uncertainty-aware strategy and the model mean with fixed offset, we selected the uncertainty-aware approach for deployment. The uncertainty-aware strategy explicitly accounts for uncertainties in estimated brushing times, which can especially help in covering brushing times when prediction errors are highly variable, as can be expected for some participants in a real-world study. The uncertainty-aware strategy has also been shown to outperform fixed offset approaches when coverage is computed in an alternative manner \cite{gazi2025sigmascheduling}.

Each participant is enrolled in the study for up to 135 days. To initialize the participant-specific models, the first 7 days of data are used as a warm-start period, during which model parameters are updated but adaptive timing decisions are not evaluated. The priors for all participants, as well as the hyperparameter \(\alpha\) used in the uncertainty-aware strategy, are initialized using the dataset described in Section~\ref{Sec:dataset}, ensuring a consistent starting point for the predictive model across individuals. For this analysis, the cut-off criterion is that participants must have completed the study as of April 20, 2026, resulting in a sample of 20 participants.

We evaluate the deployed intervention timing strategy using the same coverage-based metric as in the previous subsections. As a reference, we compare the coverage achieved by the adaptive strategy to that obtained using user-provided intervention times, which correspond to the non-adaptive strategy used in the previous deployment. The results are summarized in Table~\ref{tab:coverage_by_Ca}.
\begin{itemize}
    \item {The results observed in the ongoing clinical trial are consistent with our prior evaluations conducted before deployment. In both settings, the adaptive intervention timing strategy outperforms the fixed user-provided timing for coverage tolerance values \(C \geq 3\) hours. }
    \item{In the ongoing trial, the uncertainty-aware adaptive strategy yields consistent improvements over the baseline for \(C \geq 2\), with statistically significant gains observed for \(C \geq 3\). For example, at \(C = 3\), coverage increases from \(48.84\%\) to \(58.76\%\) (\(+9.91\%\)), and this improvement grows to \(+12.92\%\) at \(C = 4\) and \(+14.23\%\) at \(C = 5\).

    Comparing across deployments, absolute coverage values differ between the ongoing trial and the previous study, reflecting differences in participant samples and study duration. However, the relative improvement of the adaptive strategy over the baseline is consistent across both settings. We note that these findings are preliminary, and a more definitive assessment will be possible as additional participant data become available upon study completion.}
\end{itemize}

\begin{table*}[t]
\centering
\scriptsize
\setlength{\tabcolsep}{3pt}
\renewcommand{\arraystretch}{0.95}

\resizebox{\textwidth}{!}{
\begin{tabular}{c cc | cc}
\toprule
\makecell[c]{\textbf{Tolerance} \\ \textbf{(\(C\))}}
& \multicolumn{2}{c}{\textbf{Ongoing Trial (N=20, 135 days)}}
& \multicolumn{2}{c}{\textbf{Previous Trial (N=69, 70 days)}} \\
\cmidrule(lr){2-3} \cmidrule(lr){4-5}
& \makecell[c] {Baseline \\ ($U_{t}$)}
& \makecell[c]{Mean + Unc \\ ($\hat{\mu}_t - \alpha \hat{\sigma}_t$)}
&  \makecell[c]{Baseline\\ ($U_{t}$)}
& \makecell[c]{Mean + Unc \\ ( $\hat{\mu}_t - \alpha \hat{\sigma}_t$ )} \\
\midrule

\textbf{1 Hour}
& $26.00 \pm 19.78$ & $19.05 \pm 11.47$
& 18.25 $\pm$ 12.88 & \textcolor{red}{21.36 $\pm$ 12.76 (+3.10)} \\

\textbf{2 Hour}
& $39.20 \pm 20.85$ & $43.65 \pm 19.27$
& $32.75 \pm 16.71$ & \textcolor{red}{$40.37 \pm 17.29 (+7.62)^{*}$}\\

\textbf{3 Hour}
& $48.84 \pm 20.12$ & \textcolor{red}{$58.76 \pm 20.24 (+9.91)^{*}$}
& 42.53 $\pm$ 20.34 & \textcolor{red}{$52.42 \pm 18.84 (+9.88)^{*}$} \\

\textbf{4 Hour}
& $56.32 \pm 21.49$ & \textcolor{red}{$69.24 \pm 15.73 (+12.92)^{*}$}
& 48.50 $\pm$ 20.47 & \textcolor{red}{$63.30 \pm 17.23 (+14.79)^{*}$} \\

\textbf{5 Hour}
& $60.77 \pm 21.10$ & \textcolor{red}{$75.00 \pm 13.71 (+14.23)^{*}$}
& 52.64 $\pm$ 20.97 & \textcolor{red}{$70.42 \pm 15.43 (+17.77)^{*}$} \\

\bottomrule
\end{tabular}
}

\caption{Comparison of coverage (\%) between the deployed adaptive intervention timing strategy and the baseline (user-provided timing) in the ongoing clinical trial. Results from the previous trial are presented as a reference to contextualize the observed performance. Parentheses indicate the mean paired difference relative to the baseline within each study, and \(^{*}\) indicates a statistically significant difference based on a paired Wilcoxon signed-rank test (\(p < 0.05\)). Results highlighted in \textcolor{red}{red} denote, for each tolerance value within each study, the intervention timing strategy that achieves the highest coverage among those that are statistically significant relative to the baseline.}
\label{tab:coverage_by_Ca}
\end{table*}

\section{Discussion}

When interventions must be scheduled in advance of a habitual behavior, relying solely on user-provided times for these behaviors can lead to misalignment as individual routines evolve over time. As routines shift, intervention delivery may no longer coincide with the actual timing of the behavior, reducing the relevance and effectiveness of the intervention.

Our work is motivated by this timing misalignment problem observed in our previously deployed digital oral health intervention trial. In this deployment, intervention prompts were delivered at participant-provided usual brushing times collected at enrollment and held fixed throughout the trial period. Analysis of the resulting deployment data revealed that a substantial fraction of interventions were delivered either after brushing had already occurred or far in advance of it, limiting their effectiveness. Importantly, this misalignment did not arise from failures in the intervention delivery mechanism itself, but from heterogeneous changes in individual's daily routines that were not captured by static, self-reported brushing times.

This type of timing misalignment is unlikely to be unique to oral health, but is likely to arise in a wide range of digital health interventions targeting habitual behaviors, where effectiveness depends on delivering interventions at appropriate times relative to the behavior. Many such interventions rely on user-provided temporal inputs collected at onboarding, such as usual meal times in applications for diabetes management or medication adherence, and typical bedtimes or wake times in sleep and circadian rhythm interventions. In these settings, interventions are often scheduled relative to reported times under the assumption that routines remain stable. However, as with brushing, meal timing and sleep patterns can shift due to work schedules, social context, health status, or lifestyle changes. When such shifts occur, interventions based on fixed user inputs may become mistimed, even if the underlying intervention logic remains sound. The oral health case study examined in this paper illustrates how this problem can emerge in real-world deployments.

To address this challenge, we consider adaptive approaches to intervention timing that adjust intervention times as new data are observed. In our framework, we estimate brushing time using participant-specific predictive models and use these estimates to determine when to deliver interventions. We study two such strategies (\textit{Mean + Offset} and \textit{Mean + Unc.}), which differ in how the intervention time is set relative to the estimated brushing time—either using a fixed offset or incorporating predictive uncertainty. Our evaluations across both observational and simulation settings consistently show that these adaptive strategies outperform fixed intervention timing based on user input. Importantly, even when user-provided times are adjusted using a fixed offset to better precede brushing, these non-adaptive strategies still underperform compared to adaptive approaches. Notably, even relatively simple predictive models, such as BLR, provide substantial improvements in timing alignment compared to both non-adaptive strategies and more complex models.

\begin{figure*}[h]
    \centering
    \begin{subfigure}[t]{0.49\textwidth}
        \centering
        \includegraphics[width=\linewidth]{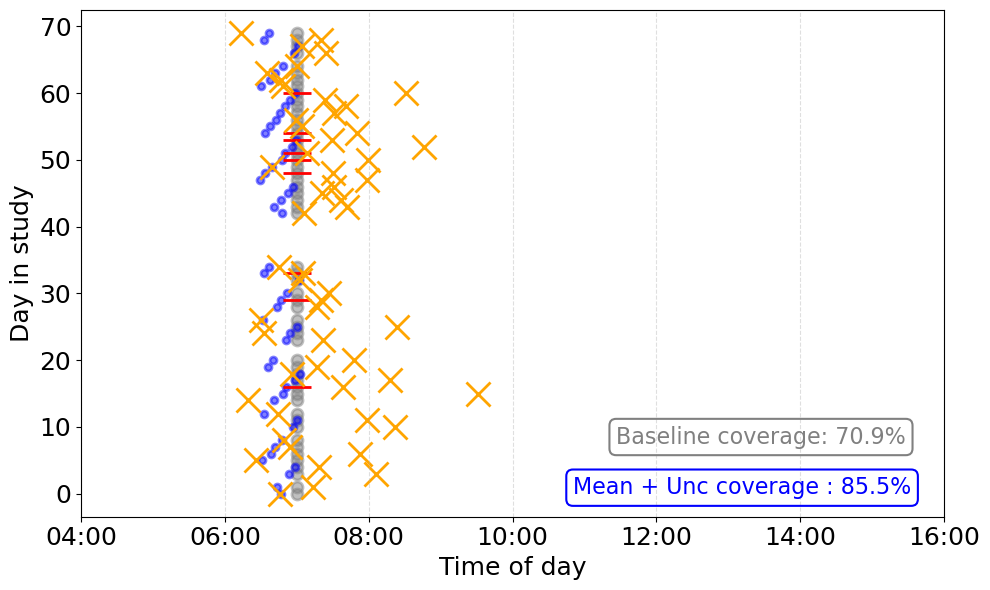}
        \caption{Participant A}
        \label{fig:participant_discussion_a}
    \end{subfigure}
    \hfill
    \begin{subfigure}[t]{0.49\textwidth}
        \centering
        \includegraphics[width=\linewidth]{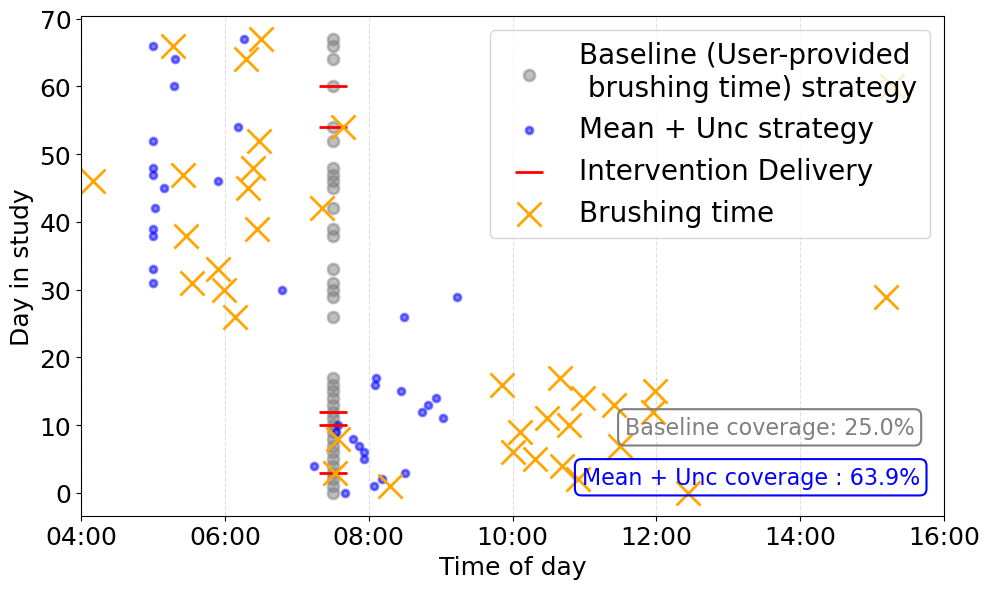}
        \caption{Participant B}
        \label{fig:participant_discussion_b}
    \end{subfigure}
    \hfill
    \begin{subfigure}[t]{0.49\textwidth}
        \centering
        \includegraphics[width=\linewidth]{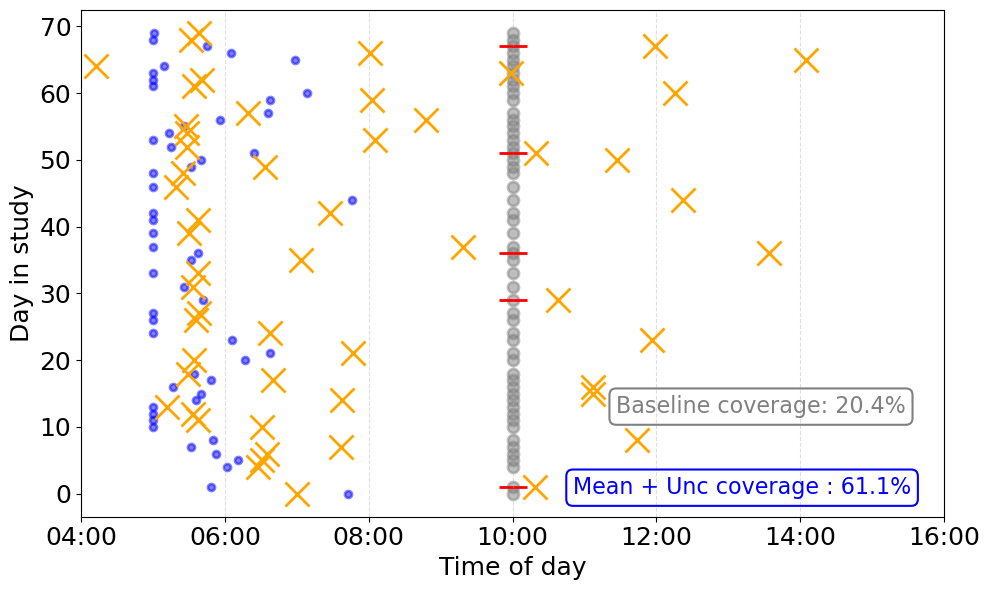}
        \caption{Participant C}
        \label{fig:participant_discussion_c}
    \end{subfigure}
    \hfill
    \begin{subfigure}[t]{0.49\textwidth}
        \centering
        \includegraphics[width=\linewidth]{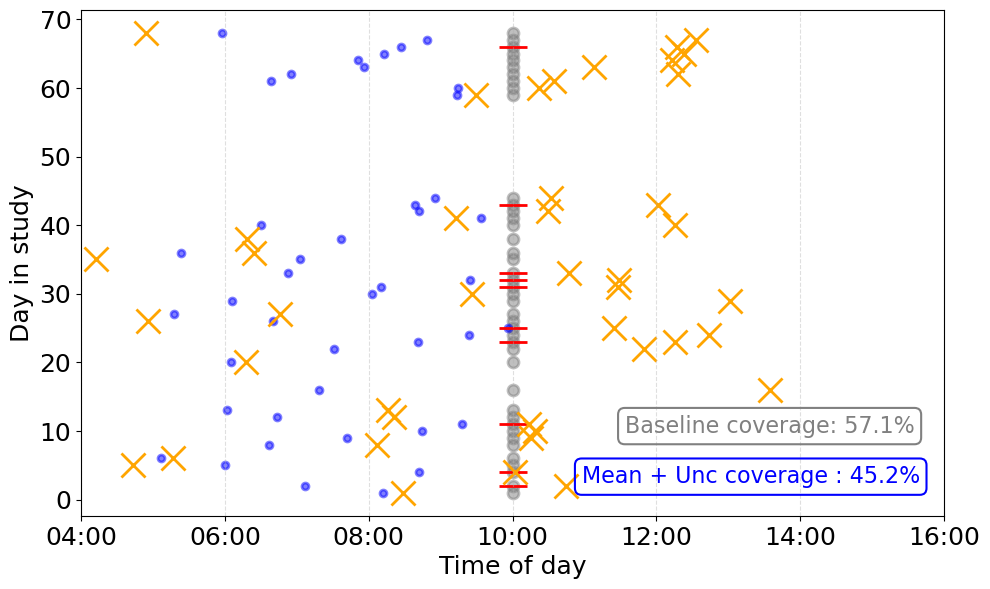}
        \caption{Participant D}
        \label{fig:participant_discussion_d}
    \end{subfigure}
\caption{Illustrative examples of participant-level brushing-time patterns and corresponding adaptive intervention times for the morning time windows over the 70-day trial period. The same participants as in Figure~\ref{fig:participant_examples} are shown, with intervention times determined using the \textit{Mean + Unc} strategy in addition to observed brushing times and user-provided usual brushing times. (a) participant whose observed brushing times remain close to their reported usual brushing time, resulting in relatively stable intervention times, (b) participant whose brushing times exhibit gradual temporal drift, with intervention times adjusting accordingly, (c) participant whose observed brushing times consistently deviate from their reported usual brushing times, with intervention times shifting earlier to improve alignment, and (d) participant whose brushing times exhibit high variability, where adaptive intervention timing performs worse than the baseline.}

    \label{fig:participant_discussion}
\end{figure*}

At the same time, these improvements are not uniform across individuals. For participants with noisy or highly variable routines, adaptive strategies may offer limited benefit. This is illustrated in Figure~\ref{fig:participant_discussion}, which presents the same participants shown in Figure~\ref{fig:participant_examples} and compares observed brushing times, user-provided times, and adaptive intervention times based on \textit{Mean + Unc} startergy. For Participant A, whose brushing times remain close to their reported usual time, adaptive intervention times show little variation and provide only modest improvement over the baseline. For Participant B, we observe a gradual shift in intervention times in response to changes in brushing times. For Participant C, whose reported times differ substantially from actual brushing times, the adaptive strategy adjusts intervention timing accordingly, moving it earlier in the day and improving alignment. In contrast, for Participant D, where brushing times exhibit high variability, adaptive strategies are unable to consistently improve over the baseline. These observations highlight that the benefit of adaptive intervention timing depends on the underlying regularity and predictability of individual routines. To better understand this heterogeneity, we conducted a post-hoc analysis using a simple regression model to examine whether improvements over the baseline were associated with participant-level demographic characteristics, including age, gender, race, and ethnicity. We did not observe any clear or consistent associations, suggesting that differences in improvement are not well explained by static demographic factors but are more likely driven by individual-specific temporal patterns.

One reason for the limited improvement in noisy cases is that our models rely primarily on features derived from historical brushing times and temporal information. Incorporating additional contextual signals—such as sleep patterns, phone usage, or daily activity rhythms—could help anticipate variability and improve robustness in such settings. In addition, the performance of participant-specific models depends on the quality of their initialization. In this work, models are initialized using population-level information via a leave-one-subject-out (LOSO) procedure, but the available covariates are relatively narrow. Access to larger population datasets or richer contextual features could enable more informative priors over timing and variability, reducing warm-up effects and improving performance, particularly for individuals with stable routines.

Building on these findings, we deployed the uncertainty-aware adaptive intervention timing strategy (\textit{Mean + Unc}) in an ongoing randomized controlled trial of the oral health digital intervention. Preliminary analysis of the deployed system is consistent with our prior evaluations, showing that adaptive intervention timing improves alignment relative to fixed, user-provided schedules, particularly for more flexible tolerance values. This consistency across observational, simulation, and deployment settings is important in light of prior work showing that predictive models based on longitudinal data can exhibit strong performance under certain evaluation settings while failing to generalize to real-world use cases \cite{langener2026just}. In contrast, our results demonstrate that the observed improvements in timing alignment persist under both retrospective evaluation and prospective deployment.

Looking forward, intervention timing can be more tightly integrated into the overall decision-making process. In this work, timing and delivery decisions are treated separately; however, future approaches could jointly optimize both \textit{when} and \textit{whether} to intervene. Additionally, adaptive timing strategies could evolve over the course of a study—for example, using user-provided schedules during an initial warm-up period and transitioning to adaptive strategies as sufficient data are collected. Another promising direction is to combine multiple timing strategies, allowing the system to dynamically select or blend strategies based on individual patterns and uncertainty.

Taken together, our results demonstrate that intervention timing is a dynamic quantity that benefits from adaptation. Treating intervention timing as a fixed design choice can lead to persistent misalignment, whereas even simple adaptive strategies can substantially improve alignment in real-world settings. These findings highlight the importance of incorporating adaptive timing into the design of longitudinal digital health interventions for habitual behaviors.

\section{Conclusion}

In this work, we addressed the problem of intervention timing in digital health interventions targeting habitual behaviors, where interventions must be scheduled in advance and fixed schedules can become misaligned as routines evolve. We proposed an adaptive intervention timing framework that continuously updates intervention times using participant-specific predictive models of brushing time and associated uncertainty. Through observational analysis, simulation-based evaluation, and preliminary results from an ongoing randomized controlled trial, we demonstrated that adaptive timing strategies consistently improve alignment compared to fixed, user-provided schedules. Our findings show that even simple models can yield substantial gains, while also highlighting the challenges posed by sparse data and variability in individual routines. Overall, this work establishes adaptive intervention timing as a key component of longitudinal digital health systems and provides a foundation for future approaches that jointly optimize \textit{when} and \textit{whether} to intervene.

\appendix

\section{Methodological Transparency and Reproducibility}
\label{sec:meta}
\subsection{Predictive Models}

We use the predictive models described in Section \ref{sec:onlinepredmodel}, including online Bayesian linear regression (BLR), online decision tree (DT), and neural network (NN) models with Monte Carlo dropout. Here, we provide additional implementation details necessary for reproducibility.

\paragraph{BLR Formulation.}
Brushing time is modeled as
\[
y_t = x_t^\top \beta + \epsilon_t, \quad \epsilon_t \sim \mathcal{N}(0, \sigma^2),
\]
with prior
\[
\beta \sim \mathcal{N}(\mu_0, \Sigma_0), \quad \sigma^2 \sim \text{Inv-Gamma}(\alpha_0, \beta_0).
\]

Given observed data \((X, y)\), the posterior parameters are
\[
\Sigma_n^{-1} = \Sigma_0^{-1} + X^\top X, \quad
\mu_n = \Sigma_n (\Sigma_0^{-1}\mu_0 + X^\top y),
\]
\[
\alpha_n = \alpha_0 + \frac{n}{2}, \quad
\beta_n = \beta_0 + \frac{1}{2}\left[(y - X\mu_n)^\top (y - X\mu_n) + (\mu_n - \mu_0)^\top \Sigma_0^{-1} (\mu_n - \mu_0)\right].
\]

The predictive distribution at input \(x\) has mean \(x^\top \mu_n\) and variance
\[
\frac{\beta_n}{\alpha_n - 1} \left(1 + x^\top \Sigma_n x \right).
\]

We model brushing time \(y_t\) as the number of hours elapsed within the corresponding time window. For example, in the morning window (starting at 04:00), a brushing event at 09:00 corresponds to \(y_t = 5\). Participant-specific BLR models are initialized using population-level priors derived from the dataset in Section~\ref{Sec:dataset}. The prior mean \(\mu_0\) is obtained by fitting a generalized estimating equation (GEE) model on pooled data across participants, retaining statistically significant coefficients and setting non-significant coefficients to zero. The prior covariance \(\Sigma_0\) is constructed using participant-level variance estimates, with variance for non-significant coefficients shrunk by a factor of 0.5. Noise prior parameters \((\alpha_0, \beta_0)\) are fixed across participants.

We construct feature vectors \(x_t\) using temporal context and historical brushing-time information observed up to the current decision window. The full set of candidate features is summarized in Table~\ref{tab:full_features}. Final model features were selected using \texttt{SelectKBest}, resulting in a reduced set of 9 features used for predictive modeling.For the BLR model, priors are constructed differently for evaluation and deployment. When evaluating on the dataset described in Section \ref{Sec:dataset}, we use a LOSO procedure to estimate participant-specific priors from data of all other participants. For the ongoing trial, priors are constructed using this full dataset. The resulting prior means and variances are summarized in Table~\ref{tab:prior_spec}.

\paragraph{Online Decision Tree (DT).}
The DT model is implemented as a Hoeffding Tree regressor and is trained incrementally using participant-specific data. The same feature construction and selection pipeline as BLR is used, including the reduced set of features obtained via \texttt{SelectKBest}. Model parameters are updated sequentially as new observations become available, and uncertainty is estimated using the variance of outcomes maintained at each leaf. The hyperparameters used for the DT model are summarized in Table~\ref{tab:dt_params}.

\paragraph{Online Neural Network (NN).}
The NN model is trained incrementally using participant-specific data and uses the full feature set without feature selection. A shared encoder is first pretrained using data from all other participants, followed by an online update of a participant-specific prediction head as new observations are collected. Predictive uncertainty is estimated using multiple stochastic forward passes under the same inference framework. Model hyperparameters are chosen via a grid search over the space shown in Table~\ref{tab:nn_params}, using validation on the dataset described in Section~\ref{Sec:dataset}.

\subsection{Simulation Environment}

For each participant in the dataset described in Section~\ref{Sec:dataset}, we construct participant-specific outcome-generating models for brushing time and app engagement.

\paragraph{Brushing Time Generation.}
For each participant in the dataset described in Section~\ref{Sec:dataset}, we fit a participant-specific survival model to characterize brushing time within each decision window \(t\). Brushing time is treated as a time-to-event outcome over a 12-hour window, with right censoring for windows in which no brushing event is observed. Let \(\tau \in [0,12]\) denote the elapsed time since the start of window \(t\). The fitted survival function \(S(\tau)\) represents the probability that brushing has not yet occurred by elapsed time \(\tau\). Brushing times are generated by sampling from this survival distribution using inverse transform sampling.

To evaluate the effect of intervention timing, we consider three variants. In the data-driven variant, brushing times are sampled directly from the fitted survival model without modification. In the adaptive variants, when an intervention is delivered (i.e., \(A_t = 1\)), the survival function is modified within a window of length \(C\) hours following the intervention time \(I_t\) to increase the probability of brushing in that interval. Specifically, we construct a modified survival function \(S'(\tau)\) such that the conditional probability of brushing within the interval \([I_t, I_t + C]\) satisfies
\[
P(\tau \leq I_t + C \mid \tau \geq I_t) = p_{\text{target}},
\]
while preserving the original survival structure outside this interval.

In the moderate-effect variant, we set \(p_{\text{target}} = 0.5\), and in the strong-effect variant, we set \(p_{\text{target}} = 1\), resulting in increasingly aggressive shifts of brushing times toward the intervention time. If no intervention is delivered, brushing times are sampled from the original survival model.

\paragraph{App Engagement.}
App engagement is modeled as a binary outcome indicating whether the participant opens the app within a time window. For each participant, we estimate the empirical probability of engagement based on the proportion of time windows in which the app was opened during the observed study period. During simulation, engagement is generated independently for each time window by sampling from a Bernoulli distribution with this participant-specific probability.

\paragraph{Simulation Validation.}
To assess the fidelity of the outcome-generating models, we compare summary statistics of simulated brushing times with those observed in the original dataset. Table~\ref{tab:simulation_brushtimes} reports aggregate statistics across participants, including censoring rates and distributional properties of brushing times (mean, median, and percentiles).

Across all variants, the simulated distributions closely match those observed in the real data, indicating that the survival-based generative models capture key characteristics of brushing-time variability. Differences across variants reflect the intended modifications to the brushing-time distribution under different assumptions about the effect of intervention timing.

\begin{table}[h]
\centering
\small
\begin{tabular}{p{4.2cm} p{8.8cm}}
\hline
\textbf{Feature Group} & \textbf{Description} \\
\hline
Decision window & Indicator for morning or evening time window. \\

Day in study & Number of days since participant enrollment. \\

Day of week & Integer-coded day of week for the decision date. \\

Current-window historical statistics & Over the previous 7 days for the same window type as the current decision window: earliest brushing time, latest brushing time, exponentially weighted average brushing time, coefficient of variation, and number of missed brushing windows. \\

Opposite-window historical statistics & Over the previous 7 days for the opposite window type: earliest brushing time, latest brushing time, exponentially weighted average brushing time, coefficient of variation, and number of missed brushing windows. \\

Most recent brushing time & Time of the most recent observed brushing event in the current window type, represented relative to the start of the decision window. \\

User-provided brushing time & Participant-reported brushing time for the current window (morning/evening, weekday/weekend), represented relative to the corresponding decision window start. \\

App engagement & Indicator for whether the participant opened the app in the previous 24 hours. \\

Polynomial feature expansion & All second-degree polynomial combinations of the above base features were additionally considered during feature selection. \\
\hline
\end{tabular}
\caption{Initial full set of features considered for predictive modeling before feature selection. Final selected features used for BLR and DT are reported in Table~\ref{tab:features}.}
\label{tab:full_features}
\end{table}

\begin{table}
\centering
\small
\setlength{\tabcolsep}{4pt}
\renewcommand{\arraystretch}{1.08}

\begin{tabular}{>{\centering\arraybackslash}p{0.12\linewidth}
                >{\centering\arraybackslash}p{0.18\linewidth}
                >{\centering\arraybackslash}p{0.23\linewidth}
                p{0.37\linewidth}}
\toprule
\textbf{Feature Index} & \(\mu_0\) \textbf{(Prior Mean)} & \(\Sigma_{0,jj}\) \textbf{(Prior Variance)} & \textbf{Feature Name} \\
\midrule
0 & -0.19 & \(1.20^2\) & Time of Day \\
1 & 0 & \(0.14^2\) & Day of Week \\
2 & 0.37 & \(0.44^2\) &  Recent Brushing Time (same window) \\
3 & -0.48 & \(1.12^2\) & Earliest Brushing Time (past 7 days, same window)  \\
4 & 0 & \(0.77^2\) & Latest Brushing Time (past 7 days, same window)  \\
5 & 1.20 & \(1.37^2\) & Exponential Avg Brushing Time (past 7 days, same window) \\
6 & -0.55 & \(1.37^2\) & Coefficient of Variation (past 7 days, same window)  \\
7 & 0.31 & \(0.50^2\) & Zero brushing count (past 7 days, same window)  \\

8 & 0.29 & \(1.69^2\) & User Provided Time \\
9 & 5.34 & \(1.80^2\) & Intercept \\
\bottomrule
\end{tabular}

\vspace{0.5em}
\[
\sigma^2 \sim \text{Inv-Gamma}(5.05,\; 22.63)
\]

\caption{Prior mean and variance specification for BLR features computed from deployed trial data .}
\label{tab:prior_spec}
\end{table}

\begin{table}
\centering
\small
\begin{tabular}{l c}
\toprule
\textbf{Parameter} & \textbf{Value} \\
\midrule
Split criterion & Variance reduction \\
Grace period & 50 \\
Max depth & None \\
Leaf prediction & Mean with variance estimate \\
\bottomrule
\end{tabular}
\caption{Hyperparameters used for the online decision tree (DT) model.}
\label{tab:dt_params}
\end{table}

\begin{table}
\centering
\small
\setlength{\tabcolsep}{6pt}
\renewcommand{\arraystretch}{1.05}

\begin{tabular}{l c}
\toprule
\textbf{Parameter} & \textbf{Values Searched} \\
\midrule
Hidden dimension (\(d_{\text{hidden}}\)) & 64 \\
Feature dimension (\(d_{\text{feat}}\)) & 32 \\
Dropout rate & \{0, 0.1\} \\
Pretraining epochs & \{10, 20, 50\} \\
Pretraining learning rate & \(\{10^{-2}, 3\times10^{-3}, 10^{-3}\}\) \\
Online head learning rate & \(\{10^{-2}, 3\times10^{-3}, 5\times10^{-3}\}\) \\
Replay buffer size & \{50, 100\} \\
Steps per online update & \{1, 3\} \\
MC dropout passes & 10 \\
\bottomrule
\end{tabular}

\caption{Hyperparameter search space used for the online neural network (NN) model.}
\label{tab:nn_params}
\end{table}

\begin{table}
\centering
\scriptsize
\setlength{\tabcolsep}{3pt}
\renewcommand{\arraystretch}{0.95}

\begin{tabular}{|p{0.31\linewidth}|p{0.13\linewidth}|>{\centering\arraybackslash}p{0.12\linewidth}|>{\centering\arraybackslash}p{0.12\linewidth}|>{\centering\arraybackslash}p{0.12\linewidth}|>{\centering\arraybackslash}p{0.12\linewidth}|}
\hline
\textbf{Metric} & \textbf{Summary} & \textbf{Data-driven} & \textbf{Strong-effect} & \textbf{Moderate-effect} & \textbf{Observed Data} \\ \hline
\rule{0pt}{10pt}

\multirow{2}{*}{\shortstack[l]{Proportion of windows \\ right-censored}}
 & Mean  & 0.44 (0.007) & 0.39 (0.006) & 0.40 (0.005) & 0.43 \\ \cline{2-6}
 & Std   & 0.24 (0.005) & 0.24 (0.007) & 0.23 (0.007) & 0.23 \\ \hline
\rule{0pt}{10pt}

\multirow{2}{*}{\shortstack[l]{Average brushing time \\ (non-censored, hours)}}
 & Mean  & 5.34 (0.05) & 4.84 (0.03) & 5.12 (0.07) & 5.48 \\ \cline{2-6}
 & Std   & 1.22 (0.04) & 1.22 (0.03) & 1.17 (0.05) & 1.13 \\ \hline
\rule{0pt}{10pt}

\multirow{2}{*}{\shortstack[l]{Median brushing time \\ (non-censored, hours)}}
 & Mean  & 5.22 (0.07) & 4.53 (0.07) & 4.90 (0.04) & 5.36 \\ \cline{2-6}
 & Std   & 1.52 (0.07) & 1.53 (0.04) & 1.48 (0.08) & 1.38 \\ \hline
\rule{0pt}{10pt}

\multirow{2}{*}{\shortstack[l]{75th percentile brushing time \\ (non-censored, hours)}}
 & Mean  & 7.09 (0.08) & 6.41 (0.04) & 6.76 (0.06) & 7.25 \\ \cline{2-6}
 & Std   & 1.79 (0.05) & 1.73 (0.05) & 1.67 (0.06) & 1.70 \\ \hline
\rule{0pt}{10pt}

\multirow{2}{*}{\shortstack[l]{90th percentile brushing time \\ (non-censored, hours)}}
 & Mean  & 8.44 (0.08) & 7.95 (0.07) & 8.23 (0.06) & 8.67 \\ \cline{2-6}
 & Std   & 1.82 (0.06) & 1.91 (0.07) & 1.78 (0.06) & 1.70 (0.04) \\ \hline
\rule{0pt}{10pt}

\multirow{2}{*}{\shortstack[l]{Standard deviation of brushing time \\ (non-censored, hours)}}
 & Mean  & 2.43 (0.02) & 2.31 (0.03) & 2.36 (0.04) & 2.48 \\ \cline{2-6}
 & Std   & 0.83 (0.03) & 0.79 (0.02) & 0.78 (0.02) & 0.79 (0.03) \\ \hline

\end{tabular}

\caption{Comparison of brushing-time statistics between simulated data and observed data. Values for simulated variants are reported as mean (standard deviation) across 100 simulation runs. The data-driven, moderate-effect, and strong-effect variants correspond to the simulation settings described in Section~\ref{sec:simulation}. All brushing-time values are expressed in hours relative to the start of the corresponding time window. For example, a value of 5.34 in the morning window corresponds to approximately 09:20 AM (i.e., 5.34 hours after 04:00).}
\label{tab:simulation_brushtimes}
\end{table}
\clearpage

\bibliographystyle{ACM-Reference-Format}
\bibliography{sample-base}

@article{vervloet2012effectiveness,
  title={The effectiveness of interventions using electronic reminders to improve adherence to chronic medication: a systematic review of the literature},
  author={Vervloet, Marcia and Linn, Annemiek J and van Weert, Julia CM and De Bakker, Dinny H and Bouvy, Marcel L and Van Dijk, Liset},
  journal={Journal of the American Medical Informatics Association},
  volume={19},
  number={5},
  pages={696--704},
  year={2012},
  publisher={BMJ Group}
}

@article{ong2023randomized,
  title={A randomized-controlled trial of a digital, small incentive-based intervention for working adults with short sleep},
  author={Ong, Ju Lynn and Massar, Stijn AA and Lau, TeYang and Ng, Ben KL and Chan, Lit Fai and Koek, Daphne and Cheong, Karen and Chee, Michael WL},
  journal={Sleep},
  volume={46},
  number={5},
  pages={zsac315},
  year={2023},
  publisher={Oxford University Press US}
}

@misc{fullagar2016time,
  title={Time to wake up: individualising the approach to sleep promotion interventions},
  author={Fullagar, Hugh HK and Bartlett, Jonathan D},
  journal={British Journal of Sports Medicine},
  volume={50},
  number={3},
  pages={143--144},
  year={2016},
  publisher={BMJ Publishing Group Ltd and British Association of Sport and Exercise Medicine}
}

@article{scheerman2020effect,
  title={The effect of using a mobile application (“WhiteTeeth”) on improving oral hygiene: A randomized controlled trial},
  author={Scheerman, Janneke FM and van Meijel, Berno and van Empelen, Pepijn and Verrips, Gijsbert HW and van Loveren, Cor and Twisk, Jos WR and Pakpour, Amir H and van den Braak, Matheus CT and Kramer, Gem JC},
  journal={International journal of dental hygiene},
  volume={18},
  number={1},
  pages={73--83},
  year={2020},
  publisher={Wiley Online Library}
}

@article{thakkar2016mobile,
  title={Mobile telephone text messaging for medication adherence in chronic disease: a meta-analysis},
  author={Thakkar, Jay and Kurup, Rahul and Laba, Tracey-Lea and Santo, Karla and Thiagalingam, Aravinda and Rodgers, Anthony and Woodward, Mark and Redfern, Julie and Chow, Clara K},
  journal={JAMA internal medicine},
  volume={176},
  number={3},
  pages={340--349},
  year={2016}
}

@inproceedings{thomaz2015practical,
  title={A practical approach for recognizing eating moments with wrist-mounted inertial sensing},
  author={Thomaz, Edison and Essa, Irfan and Abowd, Gregory D},
  booktitle={Proceedings of the 2015 ACM international joint conference on pervasive and ubiquitous computing},
  pages={1029--1040},
  year={2015}
}

@inproceedings{gal2016dropout,
  title={Dropout as a bayesian approximation: Representing model uncertainty in deep learning},
  author={Gal, Yarin and Ghahramani, Zoubin},
  booktitle={international conference on machine learning},
  pages={1050--1059},
  year={2016},
  organization={PMLR}
}

@article{yuan2024self,
  title={Self-supervised learning for human activity recognition using 700,000 person-days of wearable data},
  author={Yuan, Hang and Chan, Shing and Creagh, Andrew P and Tong, Catherine and Acquah, Aidan and Clifton, David A and Doherty, Aiden},
  journal={NPJ digital medicine},
  volume={7},
  number={1},
  pages={91},
  year={2024},
  publisher={Nature Publishing Group UK London}
}

@article{huang2025ibrush,
  title={iBrush: Toothbrushing Monitoring using Smartwatch},
  author={Huang, Hua and Lin, Shan},
  journal={ACM Transactions on Computing for Healthcare},
  year={2025},
  publisher={ACM New York, NY}
}

@article{gullapalli2021opitrack,
  title={Opitrack: a wearable-based clinical opioid use tracker with temporal convolutional attention networks},
  author={Gullapalli, Bhanu Teja and Carreiro, Stephanie and Chapman, Brittany P and Ganesan, Deepak and Sjoquist, Jan and Rahman, Tauhidur},
  journal={Proceedings of the ACM on interactive, mobile, wearable and ubiquitous technologies},
  volume={5},
  number={3},
  pages={1--29},
  year={2021},
  publisher={ACM New York, NY, USA}
}

@article{liao2020personalized,
  title={Personalized heartsteps: A reinforcement learning algorithm for optimizing physical activity},
  author={Liao, Peng and Greenewald, Kristjan and Klasnja, Predrag and Murphy, Susan},
  journal={Proceedings of the ACM on interactive, mobile, wearable and ubiquitous technologies},
  volume={4},
  number={1},
  pages={1--22},
  year={2020},
  publisher={ACM New York, NY, USA}
}

@inproceedings{parate2014risq,
  title={Risq: Recognizing smoking gestures with inertial sensors on a wristband},
  author={Parate, Abhinav and Chiu, Meng-Chieh and Chadowitz, Chaniel and Ganesan, Deepak and Kalogerakis, Evangelos},
  booktitle={Proceedings of the 12th annual international conference on Mobile systems, applications, and services},
  pages={149--161},
  year={2014}
}

@inproceedings{gazi2025sigmascheduling,
  title={SigmaScheduling: Uncertainty-Informed Scheduling of Decision Points for Intelligent Mobile Health Interventions},
  author={Gazi, Asim H and Gullapalli, Bhanu Teja and Gao, Daiqi and Marlin, Benjamin M and Shetty, Vivek and Murphy, Susan A},
  booktitle={2025 IEEE 21st International Conference on Body Sensor Networks (BSN)},
  pages={1--4},
  year={2025},
  organization={IEEE}
}

@article{nahum2016just,
  title={Just-in-time adaptive interventions (JITAIs) in mobile health: key components and design principles for ongoing health behavior support},
  author={Nahum-Shani, Inbal and Smith, Shawna N and Spring, Bonnie J and Collins, Linda M and Witkiewitz, Katie and Tewari, Ambuj and Murphy, Susan A},
  journal={Annals of behavioral medicine},
  pages={1--17},
  year={2016},
  publisher={Springer}
}

@article{walton2018optimizing,
  title={Optimizing digital integrated care via micro-randomized trials},
  author={Walton, Ashley and Nahum-Shani, Inbal and Crosby, Lori and Klasnja, Predrag and Murphy, Susan},
  journal={Clinical Pharmacology \& Therapeutics},
  volume={104},
  number={1},
  pages={53--58},
  year={2018},
  publisher={Wiley Online Library}
}

@article{langener2026just,
  title={Just in Time or Just a Guess? Addressing Challenges in Validating Prediction Models Based on Longitudinal Data},
  author={Langener, Anna M and Jacobson, Nicholas C},
  journal={Advances in Methods and Practices in Psychological Science},
  volume={9},
  number={2},
  pages={25152459261418960},
  year={2026},
  publisher={SAGE Publications Sage CA: Los Angeles, CA}
}

@article{klasnja2015microrandomized,
  title={Microrandomized trials: An experimental design for developing just-in-time adaptive interventions.},
  author={Klasnja, Predrag and Hekler, Eric B and Shiffman, Saul and Boruvka, Audrey and Almirall, Daniel and Tewari, Ambuj and Murphy, Susan A},
  journal={Health Psychology},
  volume={34},
  number={S},
  pages={1220},
  year={2015},
  publisher={American Psychological Association}
}

@article{chen2020promotion,
  title={The promotion of eating behaviour change through digital interventions},
  author={Chen, Yang and Perez-Cueto, Federico JA and Giboreau, Agn{\`e}s and Mavridis, Ioannis and Hartwell, Heather},
  journal={International journal of environmental research and public health},
  volume={17},
  number={20},
  pages={7488},
  year={2020},
  publisher={MDPI}
}

@article{pape2025effects, title={Effects of a guided digital intervention on sleep and mental health outcomes in university students--a randomized controlled trial}, author={Pape, Laura Michelle and van Straten, Annemieke and Struijs, Sascha Yuri and Karch, Julian David and Spinhoven, Philip and Antypa, Niki and Caring Universities Consortium}, journal={SLEEPJ}, pages={zsaf357}, year={2025}, publisher={Oxford University Press} }

@article{trella2024oralytics,
  title={Oralytics reinforcement learning algorithm},
  author={Trella, Anna L and Zhang, Kelly W and Carpenter, Stephanie M and Elashoff, David and Greer, Zara M and Nahum-Shani, Inbal and Ruenger, Dennis and Shetty, Vivek and Murphy, Susan A},
  journal={arXiv preprint arXiv:2406.13127},
  year={2024}
}

@article{battalio2021sense2stop,
  title={Sense2Stop: a micro-randomized trial using wearable sensors to optimize a just-in-time-adaptive stress management intervention for smoking relapse prevention},
  author={Battalio, Samuel L and Conroy, David E and Dempsey, Walter and Liao, Peng and Menictas, Marianne and Murphy, Susan and Nahum-Shani, Inbal and Qian, Tianchen and Kumar, Santosh and Spring, Bonnie},
  journal={Contemporary Clinical Trials},
  volume={109},
  pages={106534},
  year={2021},
  publisher={Elsevier}
}

@article{businelle2016ecological,
  title={An ecological momentary intervention for smoking cessation: evaluation of feasibility and effectiveness},
  author={Businelle, Michael S and Ma, Ping and Kendzor, Darla E and Frank, Summer G and Vidrine, Damon J and Wetter, David W},
  journal={Journal of medical Internet research},
  volume={18},
  number={12},
  pages={e321},
  year={2016},
  publisher={JMIR Publications Toronto, Canada}
}

@inproceedings{bhandari2017non,
  title={Non-invasive sensor based automated smoking activity detection},
  author={Bhandari, Babin and Lu, JianChao and Zheng, Xi and Rajasegarar, Sutharshan and Karmakar, Chandan},
  booktitle={2017 39th Annual International Conference of the IEEE Engineering in Medicine and Biology Society (EMBC)},
  pages={845--848},
  year={2017},
  organization={IEEE}
}

@article{ben2021smartphone,
  title={A smartphone intervention for people with serious mental illness: fully remote randomized controlled trial of CORE},
  author={Ben-Zeev, Dror and Chander, Ayesha and Tauscher, Justin and Buck, Benjamin and Nepal, Subigya and Campbell, Andrew and Doron, Guy},
  journal={Journal of medical Internet research},
  volume={23},
  number={11},
  pages={e29201},
  year={2021},
  publisher={JMIR Publications Toronto, Canada}
}

@article{naughton2017delivering,
  title={Delivering “just-in-time” smoking cessation support via mobile phones: current knowledge and future directions},
  author={Naughton, Felix},
  journal={Nicotine \& Tobacco Research},
  volume={19},
  number={3},
  pages={379--383},
  year={2017},
  publisher={Oxford University Press US}
}

@article{goldstein2020refining,
  title={Refining an algorithm-powered just-in-time adaptive weight control intervention: a randomized controlled trial evaluating model performance and behavioral outcomes},
  author={Goldstein, Stephanie P and Thomas, J Graham and Foster, Gary D and Turner-McGrievy, Gabrielle and Butryn, Meghan L and Herbert, James D and Martin, Gerald J and Forman, Evan M},
  journal={Health informatics journal},
  volume={26},
  number={4},
  pages={2315--2331},
  year={2020},
  publisher={SAGE Publications Sage UK: London, England}
}

@String{Computing = "Computing" }

@String{Springer = "Springer-Verlag" }

\end{document}